\documentclass[twocolumn]{article}
\usepackage[english]{babel}
\usepackage{blindtext}

\usepackage[%
	paperwidth=8.5in,
	paperheight=11in,
	top=1in, bottom=1in,
	left=0.75in, right=0.75in,
	columnsep=0.3in
]{geometry}

\usepackage{caption}
\usepackage[hyphens]{url}

\usepackage{booktabs} 
\usepackage[utf8]{inputenc}
\usepackage{color}
\usepackage{xcolor}
\usepackage{algorithm}
\usepackage{algorithmicx}
\usepackage{algpseudocode}
\usepackage{epsfig,endnotes,multirow,multicol}
\usepackage{ifthen}
\usepackage{datetime}
\usepackage{xspace}
\usepackage{todonotes}
\usepackage{graphicx}
\usepackage{subcaption}
\usepackage[inline]{enumitem}
\usepackage{amsmath}
\newcommand\Mark[1]{\textsuperscript#1}
\date{}
\begin{document}
\title{Seek and Push: Detecting Large Traffic Aggregates in the Dataplane}
\author{Jan Ku\v{c}era\Mark{1},
        Diana Andreea Popescu\Mark{2},
        Gianni Antichi\Mark{2},\\
        Jan Ko\v{r}enek\Mark{3},
        Andrew W. Moore\Mark{2}\\
        \small{\Mark{1}CESNET, CZ \quad \Mark{2}University of Cambridge, UK \quad \Mark{3}Brno University of Technology, CZ}}

\maketitle

\begin{abstract}
High level goals such as bandwidth provisioning, accounting and network 
anomaly detection can be easily met if high-volume traffic clusters are 
detected in real time. This paper presents Elastic Trie, an alternative 
to approaches leveraging controller-dataplane architectures.

Our solution is a novel push-based network monitoring approach that allows 
detection, within the dataplane, of high-volume traffic clusters. 
Notifications from the switch to the controller can be sent only as required, 
avoiding the transmission or processing of unnecessary data. Furthermore, 
the dataplane can iteratively refine the responsible IP prefixes allowing a 
controller to receive a flexible granularity information. We report 
and discuss an evaluation of our P4-based prototype, showing our 
solution to be able to detect (with 95\% of precision), hierarchical 
heavy hitters and superspreaders using less than 8KB or 80KB of active 
memory respectively. Finally, Elastic Trie can identify changes 
in the network traffic patterns, symptomatic of Denial-of-Service attack 
events.	
%
\end{abstract}

\section{Introduction}
The importance of finding high-volume traffic clusters has been recognized in the past 
to improve network management practices~\cite{estan02,curtis11,jose11,liu16,sivaraman17}.
Specific applications include, as shown in Table~\ref{tab:measure-manage}, accounting~\cite{duffield2001,estan02}, 
traffic engineering~\cite{feldmann2001, benson11}, anomaly detection~\cite{lakhina2004,lakhina2005}, 
Distributed Denial-of-Service (DDoS), and scans detection~\cite{xie2005, venkataraman2005}. 

\begin{table}[b]\centering
    \begin{tabular}{|c|c|}\hline
        \textbf{Network event}
            & \textbf{Management task} \\\hline\hline
        \parbox[c]{2.8cm}{\vspace{0.15em}\centering (Hierarchical) \\Heavy Hitters\vspace{0.15em}}
            & accounting, traffic engineering  \\\hline
        \parbox[c]{2.8cm}{\vspace{0.15em}\centering Changes in \\traffic patterns\vspace{0.15em}}
            & anomaly detection, DoS detection \\\hline
        Superspreaders
            & worm, scan, DDoS detection \\\hline
    \end{tabular}
    \vspace{0.1in}
    \caption{Detecting high-volume traffic clusters is beneficial for a number of network management tasks.}
    \label{tab:measure-manage}
\end{table}

Dataplane monitoring is the main instrument that enables the detection of high-volume 
traffic clusters. In the past, it was based on packet sampling~\cite{NetFlow,sFlow}, 
to lower overheads and data collection bandwidth, thus impacting estimation 
accuracy~\cite{choi2002, estan2004, mai2006}. OpenFlow (OF)~\cite{mckeown2008} did 
not improve the situation either~\cite{curtis11}: the main monitoring mechanism 
exposes the per-port and per-flow counters available in the switches. An application 
running on top of the controller can periodically poll counters using the 
standard OF APIs, and then perform a software-based algorithm to get insights into 
the network behavior. However, this approach limits significantly the original 
flexibility intended by Software Defined Networking (SDN). While increasing 
the gap between two consecutive counters requests reduces the controller ability to 
react in a timely fashion, continuously requesting counters from switches leads 
to non-scalable solutions by challenging the switch-controller interactions  
capabilities~\cite{curtis11}. For this reason, lately a number of proposals suggest the
use of programmable 
dataplanes using the P4 programming language~\cite{bosshart14} to extend dataplane 
functionality with more advanced monitoring features. 
Specifically, some of them leverage dataplane programmability to either directly provide 
the top-k heavy hitters~\cite{sivaraman17} or to assist the controller by exporting 
smart representations of aggregated traffic statistics~\cite{yu2013,liu16}.
While the latter case results in a more flexible and generic approach compared 
to the former, the controller still needs to receive at a fixed time interval 
the generated information from the dataplane, and estimate the various application-level 
metrics of interest based on the received data. Such an architecture might suffer from the 
same problem of the legacy OF protocol: the ability to apply network policy updates 
based on the received data depends on the switch-controller's interactions capabilities 
of collecting statistics at short time ranges~\cite{curtis11}.

In this paper we take a different approach. We leverage dataplane programmability to 
transform the switch from a passive monitoring infrastructure to an \textit{active} system 
which is capable of detecting several types of network events associated with high-volume 
traffic clusters, and only subsequently to
inform the controller. We designed a new data structure, \emph{Elastic Trie}, 
that enables the detection of hierarchical heavy hitters, changes in network traffic and 
superspreaders from within the dataplane, and we present its implementation in P4. 
The basic idea behind the proposed solution 
is to create a hash table based prefix tree that grows or collapses to focus only on the 
prefixes that account for a "large enough" share of the traffic. This enables the detection 
of (hierarchical) heavy hitters, and by looking at its growing rate it is possible to identify 
changes in the traffic patterns.

The main contributions of the paper are as follows:
\begin{itemize}[leftmargin=0.26in]
\item We propose a \textit{push-based} approach to network monitoring, where the dataplane
      informs the control plane only when specific conditions are met.
\item We present a data structure that enables the detection of a number of network events
      associated with high-volume traffic clusters within the dataplane. Specifically, 
      we demonstrate how Elastic Trie allows to detect hierarchical heavy hitters, changes 
      in network traffic and superspreaders. Moreover, our solution iteratively refines the 
      responsible prefixes so that the controller receives a finer or coarser grained information 
      depending on the desired reporting time.
\item We implemented our idea in P4 using match-action tables and we demonstrate its 
      detection capabilities by evaluating it through trace-driven simulations.
\end{itemize}


\section{Motivating Event Triggered Monitoring}
\label{sec:motivation}
We first ran an experiment to measure the amount of time it takes to retrieve
an increasing number of hardware counters from a switch. We used two different 
switches. The first is a fairly new OpenFlow-enabled IBM solution, i.e., 
RackSwitch G8264~\cite{ibm_g8264}, capable of 1.2 Tbps throughput. The second 
is the NoviSwitch 1132~\cite{noviswitch}, which has been designed for use 
in high bandwidth and flow-intensive network deployments. We connected the switches 
to a server running an OpenFlow controller and we built an application that 
allows to request an increasing number of flow counters from the switches which 
were idle when the counters were pulled.

\begin{figure}[tb]
        \centering
        \includegraphics[width=1\columnwidth,trim={7 2 1.5 8.5},clip]{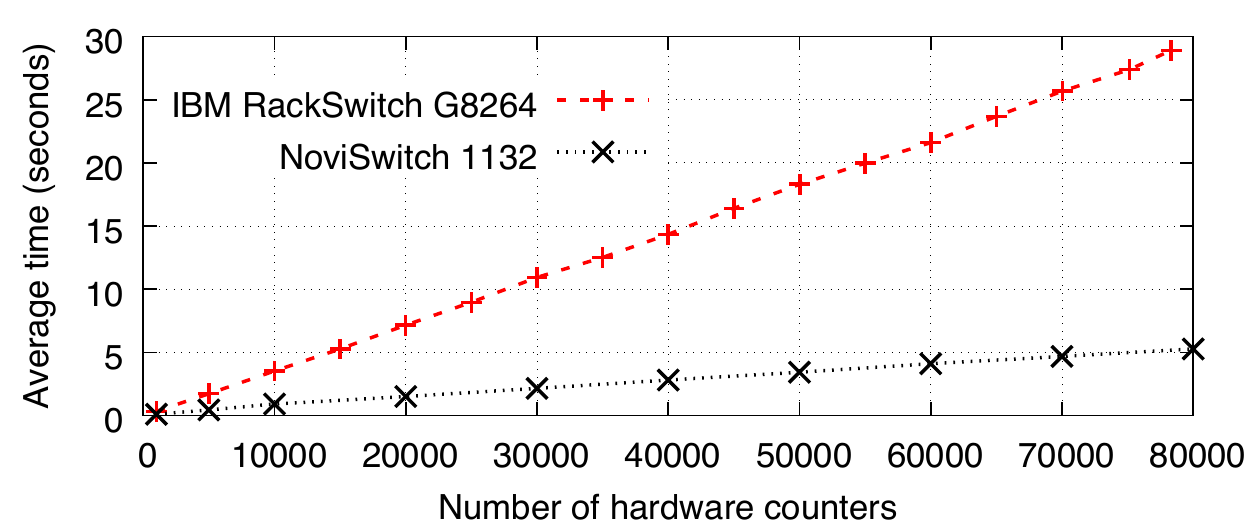}
    \vspace{-0.25in}        
        \caption{Time to retrieve hardware counters.}
        \label{fig:stat_req}
\end{figure}

Figure~\ref{fig:stat_req} shows the results we obtained. Surprisingly, the IBM switch
reports values aligned with tests performed against much older solutions~\cite{curtis11},
while the NoviSwitch performs much better. We did not manage to perform our test for 
more than 100K counters, but the increasing trend of the figure for both switches is 
clear. Although dataplane programmability can help in reducing the number of 
counters exported by aggregating flow rules in probabilistic data structures, such 
as bloom filters or sketches, past research has shown that around 150K counters are 
still required to provide useful information to the controller~\cite{liu16}. While 
the IBM switch can take up to 28 seconds to retrieve approximately half of the aforementioned 
amount of counters, the NoviSwitch needs at least 5 seconds. Certainly, such delays are not acceptable 
when it comes to critical network events detection, e.g., DoS, DDoS, scans, worms.
The lesson learned is that retrieving a large number of flow counters from hardware 
is time consuming.

\begin{figure}[tb]\centering
    \includegraphics[width=1\columnwidth,trim={7 2 1.5 8.5},clip]{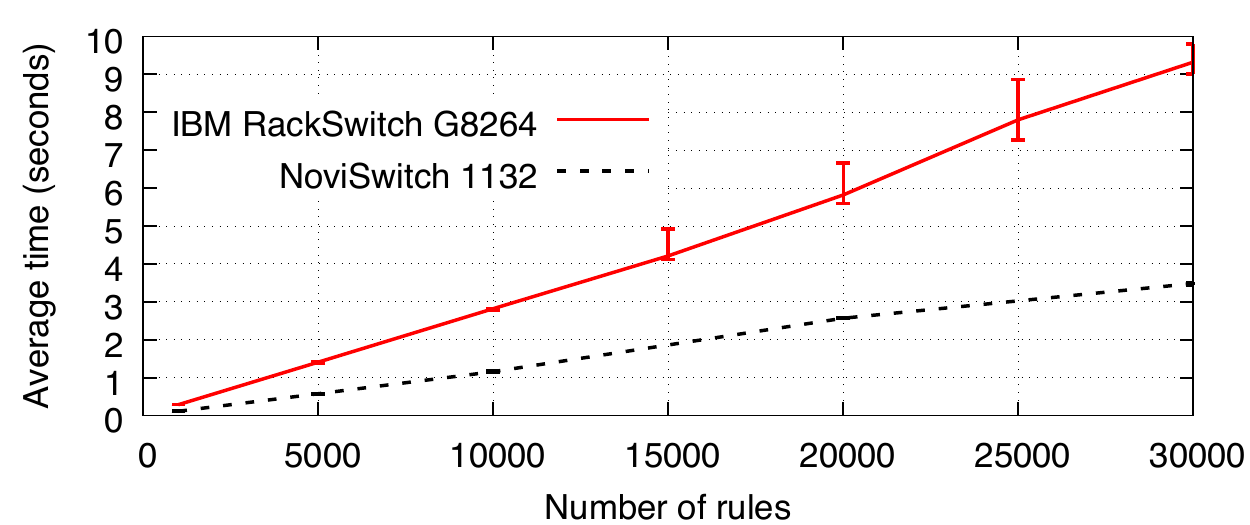}
    \vspace{-0.27in}
    \caption{Time to add new rules.}
    \label{fig:flow_add}
\end{figure}

Using the same configuration as in the previous experiment, we ran the second test to characterize
the amount of time it takes to modify an increasing number of rules on a high-end switch.
Updating the forwarding state and retrieving statistics from a controller are two
competing operations that are commonly performed sequentially by the switch. The larger the
number of flow additions or statistics requests, the bigger the impact of one action on the
completion of the other~\cite{curtis11}. For this reason it is important to characterize
switch rule update time, especially because issuing a large number of forwarding updates in
a single batch is a common defense practice for Internet Service Providers (ISPs) to stop DDoS
attacks, and can involve up to 50K rules update~\cite{giotsas17}.

Figure~\ref{fig:flow_add} shows the results we obtained. The reported values show a lower 
bound of the switch update time. In the test we pushed an increasing number of rule modifications 
with the same priority. This is the best case for a switch, as demonstrated by past research~\cite{he15}, 
as TCAMs do not have to reorder the hardware rules. Adding 30K rules, in the best case scenario, might
require 10 seconds for the IBM switch and almost 4 seconds for the NoviSwitch. This would affect 
the statistics collection capabilities of the controller.

The results obtained motivated us to build a solution that: (1) does not depend on statistics 
retrieval from the dataplane, and (2) decouples the monitoring and the forwarding statistics 
updates. Indeed, with Elastic Trie the controller receives a push-based notification only 
when an event related to high-volume traffic clusters has been detected in the dataplane.

\section{Desired Properties}
\label{ssec:desired_prop}
\begin{figure}[tb]
        \centering
        \includegraphics[width=0.90\columnwidth]{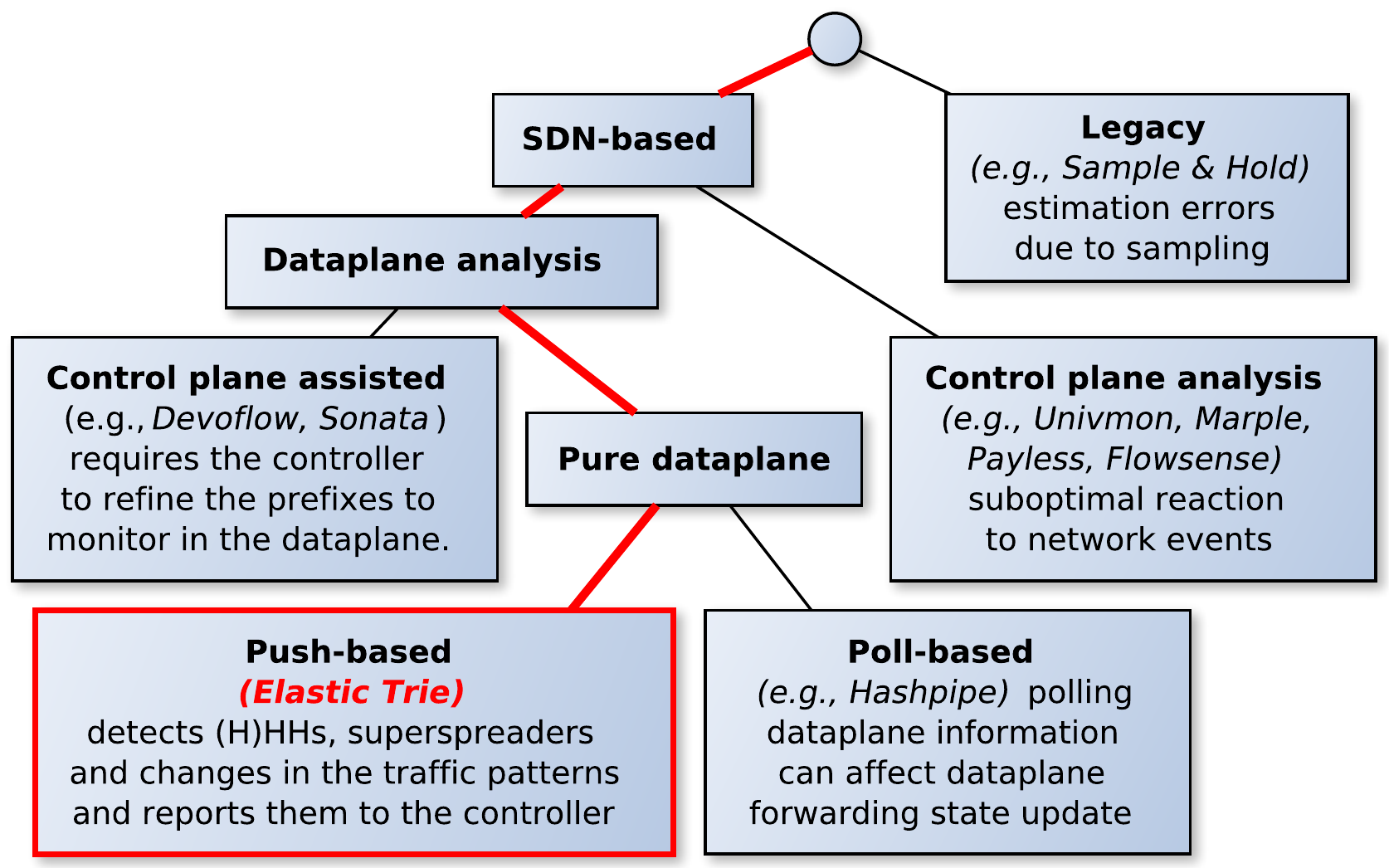}
        \vspace{-0.08in}
        \caption{Design space for detection of high volume traffic clusters.}
        \label{fig:design_space}
\end{figure}

Figure~\ref{fig:design_space} surveys the design space for the detection of high 
volume traffic clusters and places our solution, \emph{Elastic Trie}, in context by 
following the thick red lines through the design tree. 
This section describes the insights that inform our major design decisions
and provides the necessary background for the selected network events.
In designing 
Elastic Trie, we targeted a solution with a number of key features:

\textbf{Efficiency.} Collecting counter statistics from all the active 
flows or smart representations of aggregated traffic statistics~\cite{yu2013,liu16}
can create considerable control plane load. With Elastic Trie, we aim for a push-based
solution which exports information to a controller only when a network 
event has been detected in the dataplane.

\textbf{Packet processing independence.} The main OpenFlow mechanism for 
dataplane monitoring exposes the per-port and per-flow counters available in the 
switch\-es~\cite{mckeown2008}. Although this might seem a logical and simple 
solution, it suffers from a major drawback: expensive TCAM resources must be shared
between the rules needed for packet processing and the rules installed for monitoring 
purposes only~\cite{zhang13}. In addition, the use of pre-configured
monitoring rules requires prior knowledge of the active network flows, as well 
as a large number of fine grained rules, in order to accurately
detect heavy flows. With Elastic Trie, we aim to decouple the packet 
processing logic from the monitoring mechanism. While the former can still use 
the available TCAM resources, the latter can be implemented algorithmically 
with match-action tables using the P4 programming language.

\textbf{Historical network trend awareness.} Change 
detection is the process of identifying flows that contribute the most to traffic 
change over two consecutive time intervals~\cite{callegari12}. Previous solutions~\cite{liu16,sonata}
rely on the controller to compute the differences from multiple intervals,
effectively slowing down the reaction capability of the network if an anomaly
has been found. With Elastic Trie, we seek a solution capable to directly 
compute such an operation within the dataplane at the expense of minimal memory
consumption.

\textbf{Optimization for fast traffic steering.} Networks today rely on 
middleboxes to provide security and added-value services~\cite{sherry12}. 
Taking advantage of global network knowledge of a controller, it is easy to 
enforce a network policy change and steer (part of) the traffic if an anomaly
has been detected~\cite{qazi13,he17}. Collecting statistics from the dataplane, 
running the detection algorithm in the control plane, and then enforcing a policy 
change if something suspicious is found can cause delays that might not be acceptable 
in the case of a network attack. With Elastic Trie, we propose to detect at short 
timescales a coarse-grained approximation of the prefix responsible for the network 
traffic changes. Once the detection of the anomalous subnet is done in the dataplane, 
the traffic can be instantaneously redirected to the appropriate middlebox, without 
the need to communicate with the controller.

\textbf{Optimization for network management.} Dividing the time in fixed 
intervals can simplify the detection of a number of network events, e.g., heavy 
hitter, superspreader, DDoS. At the end of each time window, it is possible 
to identify the flows that consume more than a fraction $T$ of the link capacity, 
i.e., heavy hitter, or determine the host that contacts more than a number of unique 
destinations, i.e., superspreader. For this reason, current solutions for network 
monitoring typically operate by exporting counters or specific data structures, e.g., 
sketches, to the controller at fixed time scales~\cite{yu13,liu16,li16}. However, this 
approach tightly bounds the reactive capabilities of the network with the dataplane 
statistics reporting time, as it needs to be (at least) comparable to traffic 
variations~\cite{alfares10,benson11}. Only if this last condition is met, 
solutions like dynamic routing of heavy flows~\cite{curtis11,benson11,rasley14} or 
dynamic flow scheduling~\cite{sivaraman16} can be easily implemented. 
However, state of the art solutions adopt a fairly large reporting time (typically
20 seconds~\cite{liu16,sivaraman17}) not to overload the controller with too much 
data, thus limiting network reaction capabilities. 
With Elastic Trie, we target a solution that iteratively refines the responsible 
prefixes in the dataplane. In this way, the controller, depending on the desired 
reporting time, can receive finer or coarser grained information on the flow 
responsible for a network event associated with a high-volume traffic cluster.

\subsection{Selected Network Events}
\label{sec:problem_definition}
This section provides the necessary background for the network events 
considered in this paper.

\begin{figure}[t]\centering
    \includegraphics[width=8.3cm]{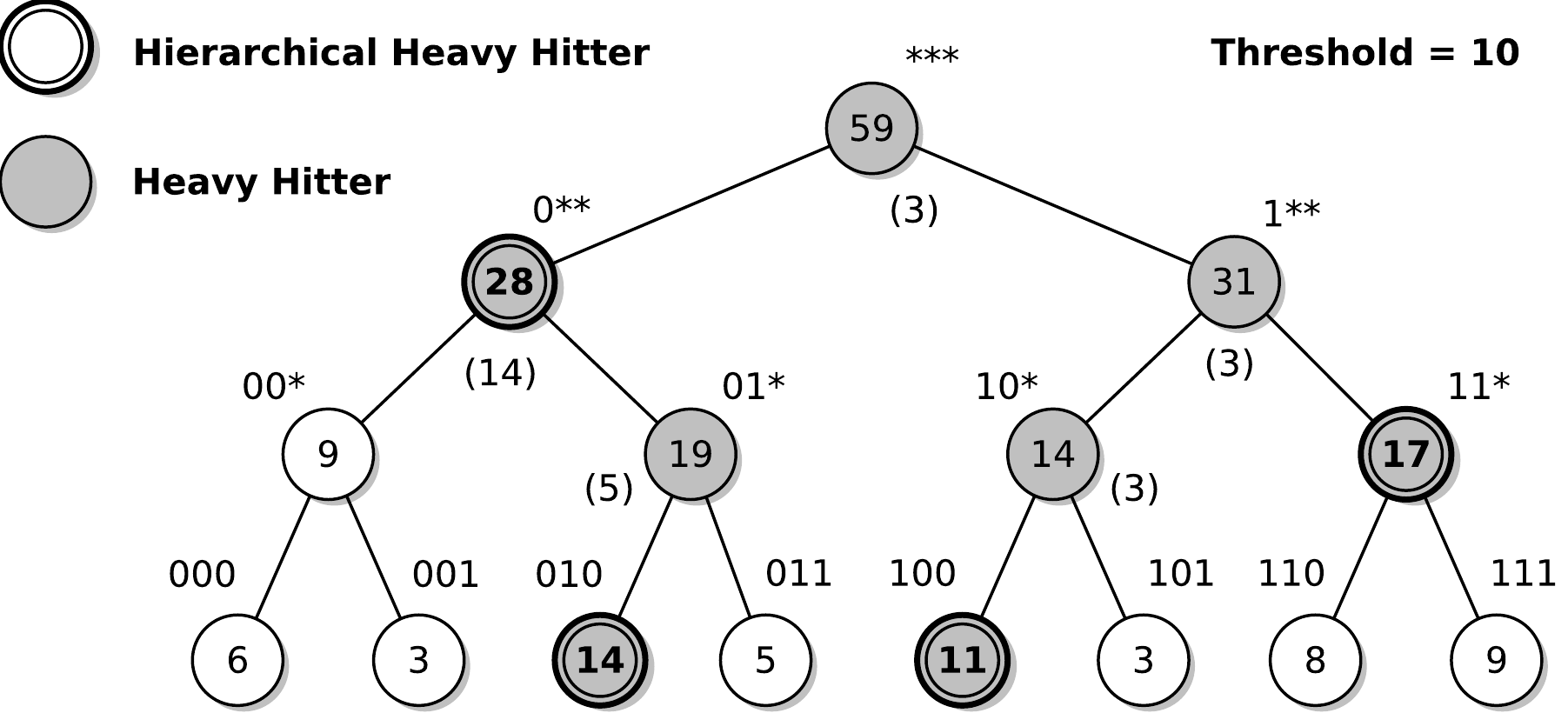}
    \vspace{-0.08in}
    \caption{A trie of IP addresses in reduced 3-bit model. Each node represents a prefix $p$ with associated amount of traffic sent. Assuming threshold $T=10$, grey nodes are heavy hitters, while double circle nodes are also hierarchical heavy hitters.}
    \label{fig:hh-vs-hhh}
\end{figure}

\textbf{(H)HH detection.}
Hierarchical Heavy Hitters have already been studied in a number of prior
works~\cite{zhang04, cormode08, jose11, mitzenmacher2012, cho2017, basat2018}.
Detecting an Heavy Hitter (HH) means identifying a large aggregate 
in the network traffic. For example, assuming the use 
of the source IP address as a key, the goal of the HH detection problem is to find 
the source IP prefixes that contribute with a traffic volume\footnote{It can be 
considered in terms of packets or bytes per second.} larger than a 
given threshold $T$ during a fixed time interval $t$.
Figure~\ref{fig:hh-vs-hhh} depicts the amount of traffic for prefixes in a reduced 
3-bit wide model domain of IP addresses. All of the prefixes are denoted as a prefix 
tree, also known as trie. Each node of the trie has at most two children. The left child 
is associated with bit value $0$, while the right child is associated with bit value $1$, 
and the prefix $p$ represented by a node is defined by the path from the root to that node. 
Terminal nodes express only the traffic volume produced by full IP addresses. Non-terminal 
nodes then summarize the traffic of a prefix. The contribution of each prefix is represented 
as a number in each node. Considering the use of a threshold $T=10$, terminal nodes \texttt{010}, 
\texttt{100}, non-terminal node \texttt{11*} and all their ancestors are identified as 
heavy hitters. For example, each child of the \texttt{11*} node contributes independently less than 
the threshold $T$, but in total both children contribute enough to exceed the threshold 
and report the \texttt{11*} prefix as a HH.

A Hierarchical Heavy Hitter (HHH)~\cite{cormode08} is a special case of HH. Specifically,
it is a prefix $p$, which exceeds a threshold $T$ after excluding the contribution 
of all its HHH descendants\footnote{The descendant prefixes need to satisfy the 
definition of HHH.}.
In Figure~\ref{fig:hh-vs-hhh}, only prefixes \texttt{010}, \texttt{100}, \texttt{0**} and 
\texttt{11*} are HHHs. The amount of traffic of each HH prefix without the impact of its 
HHH descendants is shown in brackets. In this example, the \texttt{11*} node is an HHH, as none of its children 
contributes enough to exceed the threshold $T$, but the amount of traffic from both children exceeds the threshold. 
In contrast, the \texttt{1**} prefix is not an HHH because a significant part of its contribution 
originates from its descendant nodes \texttt{100} and \texttt{11*}, which are already HHH and must be excluded.

It is worth noting that, while the detection of HHHs requires the knowledge of the HHs, 
the opposite is not true. Reporting to a controller the HHHs guarantees minimum 
overhead, while providing all the necessary information.
Taking as an example the configuration of Figure~\ref{fig:hh-vs-hhh}, a dataplane capable of 
detecting HHs would export to the controller the following prefixes: \texttt{0**, 1**, 
01*, 10*, 11*, 010} and \texttt{100}. In contrast, a dataplane with HHH detection 
capability would report just \texttt{0**, 11*, 010} and \texttt{100}.
In both cases the amount of useful information for network management practices is the 
same\footnote{Some of the reported HHs are just prefixes of more specific HHs.}, but in 
the second case we export less data.

\textbf{Change detection.} 
Traffic anomalies are a normal occurrence in the daily life of network operators. While some 
of them can be sometimes tolerable, others are often an indication of performance bottlenecks due 
to flash crowds~\cite{jung02}, network element failures, or malicious activities such as Denial-of-Service 
attacks (DoS), worms and spam. Change detection is one of the main approach to network anomaly detection. 
The method detects traffic anomalies by deriving a model of normal 
behavior based on the past traffic history and looking for significant changes in short-term 
behavior that are inconsistent with the model~\cite{krishnamurthy03}. Identifying the flows 
responsible for the changes in the traffic patterns can be formulated also
(at least in part) as a high-volume traffic clusters detection problem~\cite{estan02}. Specifically, 
it requires the ability to discover which flows contribute the most to the traffic changes over two
consecutive time intervals~\cite{callegari12}.

\textbf{Superspreader detection.} 
A superspreader is defined to be a host that contacts at least a given number of distinct destinations 
over a short time period. It can be responsible for fast worm propagation, so detecting it early is of 
paramount importance~\cite{venkataraman05}. Moreover, superspreader detection can be seen as a 
high-volume traffic cluster identification problem. Specifically, while past examples, such as HH,
typically define the traffic volume in terms of packets or bytes per second, in the case of
superspreaders, the problem is tackled in the dimension of flows per second. While an HH is
a source that sends a lot of traffic, a superspreader is a source that contacts many distinct 
destinations. In addition, superspreader detection can be seen also as Distributed Denial of Service 
(DDoS) victim detection if, instead of the source, the same type of spread detection is applied to the 
destination~\cite{yu2013}.

\section{Elastic Trie Algorithm}
The Elastic Trie algorithm is inspired by past works on HHH detection~\cite{zhang04,jose11}. 
Specifically, it operates in the same hierarchical manner. It also enables the 
detection of a number of network events associated with high-volume traffic 
clusters from within the dataplane without the need to be coordinated by a 
controller. Finally, it operates in a packet-driven manner and can 
be implemented using common match-action based architectures such as RMT~\cite{bosshart13}.

In this section, for the sake of simplicity, we first introduce a basic 
version of the Elastic Trie which is capable to detect only 
HHHs. We then describe its mapping into appropriate P4 constructs, e.g., 
match-action tables and registers. Finally, we discuss extensions to the basic 
algorithm to support the detection of other events, such as superspreaders and 
network traffic changes.

\subsection{Data Structure \& Basic Algorithm}
\label{algo}
Let us assume that we know in advance all the potential HHHs in a network. 
In this case, to correctly detect which of them is a real HHH, for each new arriving 
packet it is necessary to lookup the longest prefix matching (LPM) in the table of 
potential HHHs and then increment its associated counter. Thus, in some aspects, this 
is similar to the IP lookup problem~\cite{srinivasan98}, where the longest prefix 
matching in the forwarding table is searched. In practice, the two problems, while 
sharing some common aspects, are quite different. In the first case, the forwarding 
table is computed by the control plane, does not directly depend on the nature of 
the dataplane traffic and does not change very frequently. In contrast, a table 
storing HHHs is very dynamic, as it is correlated with the properties of the dataplane 
traffic. In addition, since the HHH prefixes are not known in advance, all 
the traffic received needs to be monitored to properly build the corresponding HHHs 
table.

The nature of the HHH problem (IP addresses can be naturally organized according to 
prefixes into a hierarchy) led us to use a tree-based data structure. Thus, for the 
purpose of HHH detection we maintain a standard trie data structure~\cite{srinivasan98}. 
A trie is a tree data structure, where the position of a node in the tree defines the 
key associated with it. Every node in a trie has at most two child nodes\footnote{For 
the sake of simplicity let us now ignore the multi-bit tries.}. 
The left child is associated with bit value 0, and the right child is associated with 
bit value 1. Each node also represents a prefix, which is defined by the path from the 
root of the tree to that specific node. With Elastic Trie, we further extend this concept 
and associate a specific data structure within each node.
%
Specifically, it consists of three elements: the counter associated to the left
child (\texttt{32 bits}), the one associated to the right child (\texttt{32 bits}) and 
a timestamp (\texttt{48 bits}). The counters represent the amount of traffic, e.g., packets 
or bytes, for each of the node's direct subprefix, while the sum of the counters represents 
the amount of traffic sent by the prefix itself. The timestamp specifies the time when 
the node was created or the last time when the counters were reset. 

The starting condition is associated to a trie composed by a single node, corresponding 
with the zero-length prefix \texttt{*}. The basic idea behind the proposed solution is to have 
a trie that grows or collapses to focus on the nodes associated to prefixes that account for 
a "large enough" share of the traffic. Thus, we named our data structure Elastic Trie. To 
achieve this, inspired by the NetFlow~\cite{NetFlow} terminology, we defined two time 
intervals: \emph{active timeout $t_A$} and \emph{inactive timeout $t_I$}, where $t_A < t_I$. 
The active timeout $t_A$ is the interval after which the prefix is evaluated and possibly 
reported as HHH to the controller. The inactive timeout $t_I$ defines the interval 
after which the IP prefix corresponding to the node is considered inactive and its counters 
outdated. Figure~\ref{fig:decision-diagram} depicts key steps of the proposed Elastic Trie
algorithm. For every incoming packet, the longest prefix (thus its corresponding node) is 
looked up and the packet timestamp ($t_P$) compared against the node timestamp ($t_N$).
Let us also denote by $c_0$ and $c_1$ the left and the right child counters of the found node.
Here, there are five possible cases that have to be considered based on the result of the
comparison, the node counter values, timeouts $t_A$ and $t_I$:

\begin{figure}[t]\centering
        \includegraphics[width=1\columnwidth]{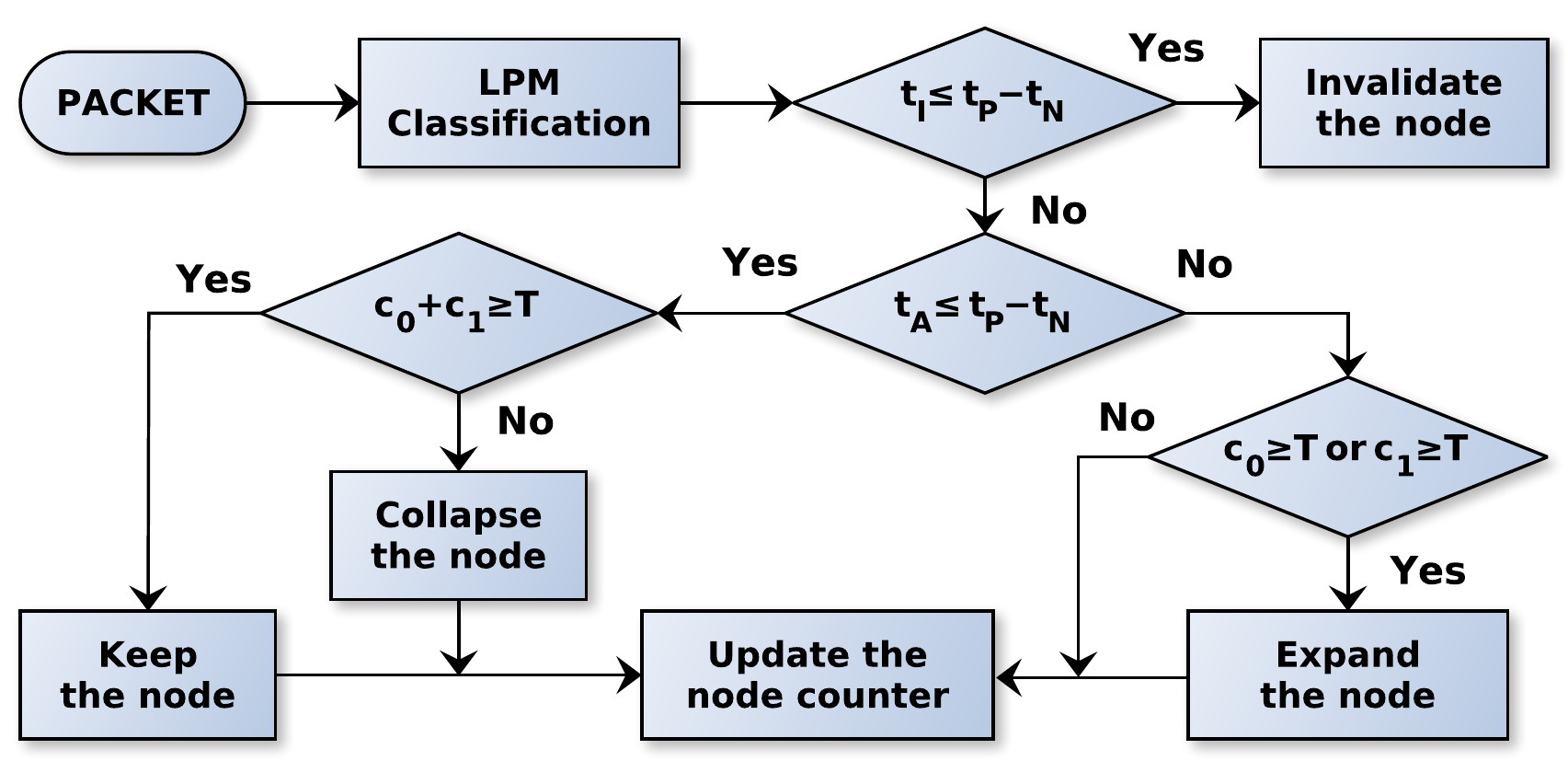}
        \vspace{-0.25in}
        \caption{Flowchart showcasing input packet processing of the
Elastic Trie detection algorithm.}
        \label{fig:decision-diagram}
\end{figure}

\begin{figure*}[t]\centering
        \begin{subfigure}[tc]{0.245\textwidth}\centering
                \includegraphics[height=3.25cm]{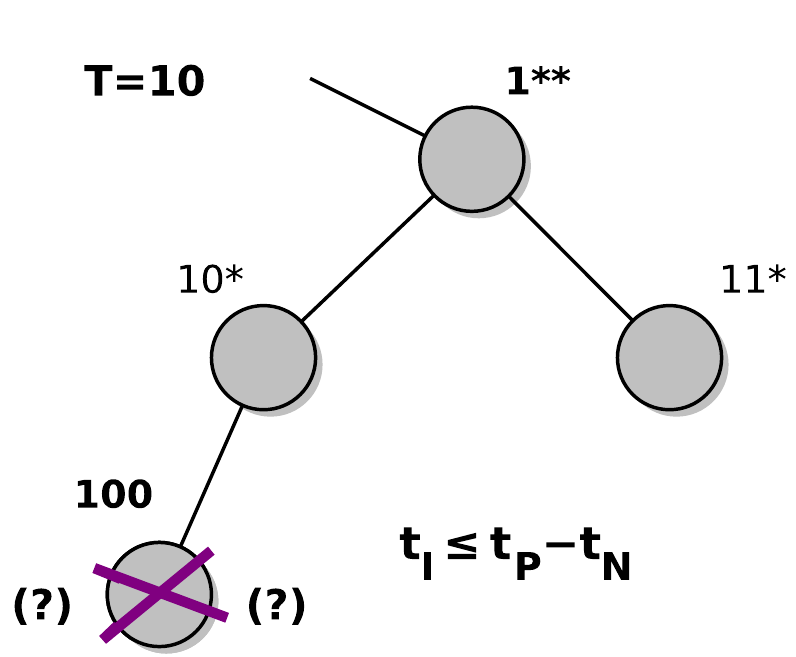}\vspace*{0.5em}
                \caption{Invalidating the node.}
                \label{fig:invalidating-node}
        \end{subfigure}  
        \begin{subfigure}[tc]{0.245\textwidth}\centering
		\includegraphics[height=3.25cm]{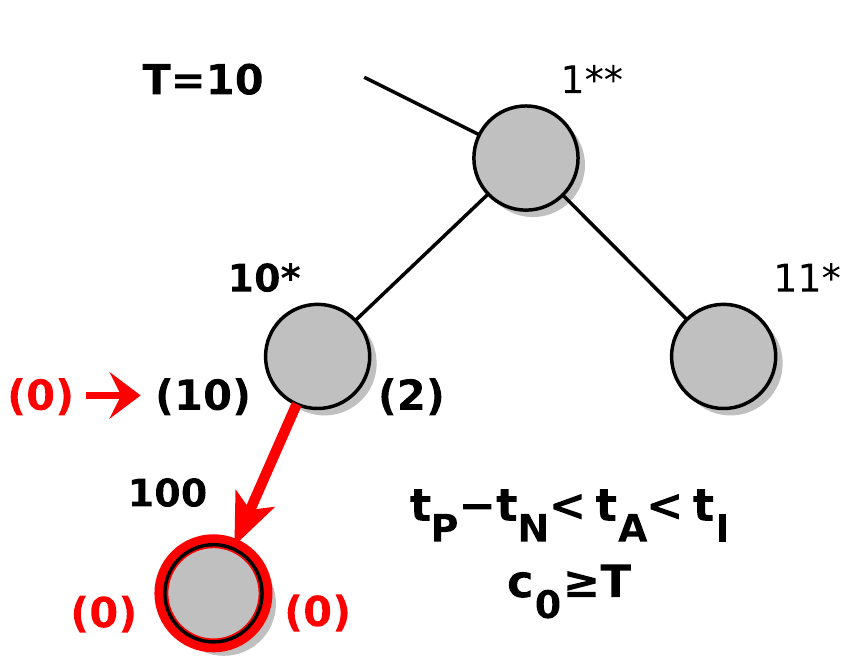}\vspace*{0.5em}
                \caption{Expanding the node.}
                \label{fig:expanding-node}
        \end{subfigure}
        \begin{subfigure}[tc]{0.245\textwidth}\centering
                \includegraphics[height=3.25cm]{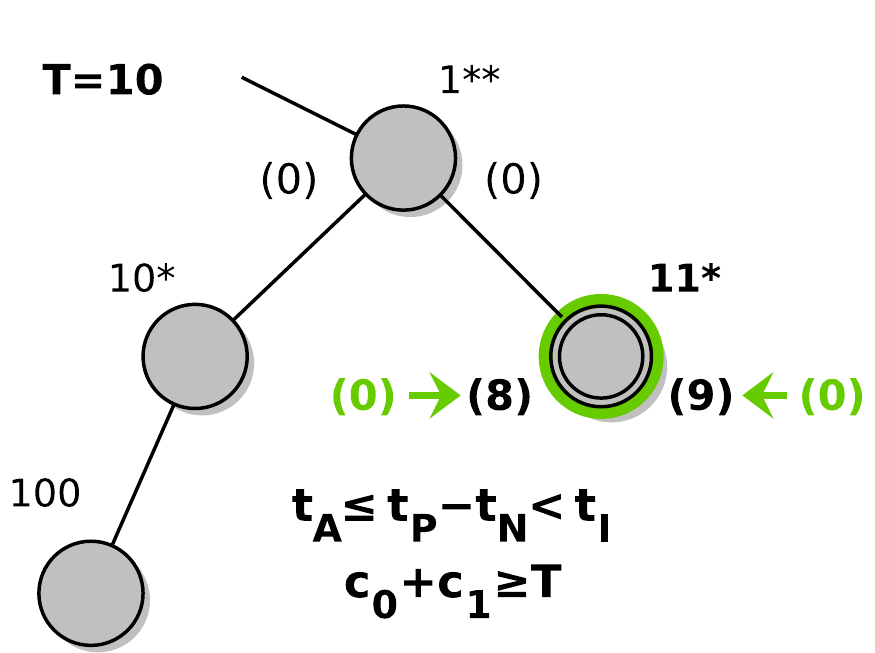}\vspace*{0.5em}
                \caption{Keeping the node.}
                \label{fig:keeping-node}
        \end{subfigure}
        \begin{subfigure}[tc]{0.245\textwidth}\centering
                \includegraphics[height=3.25cm]{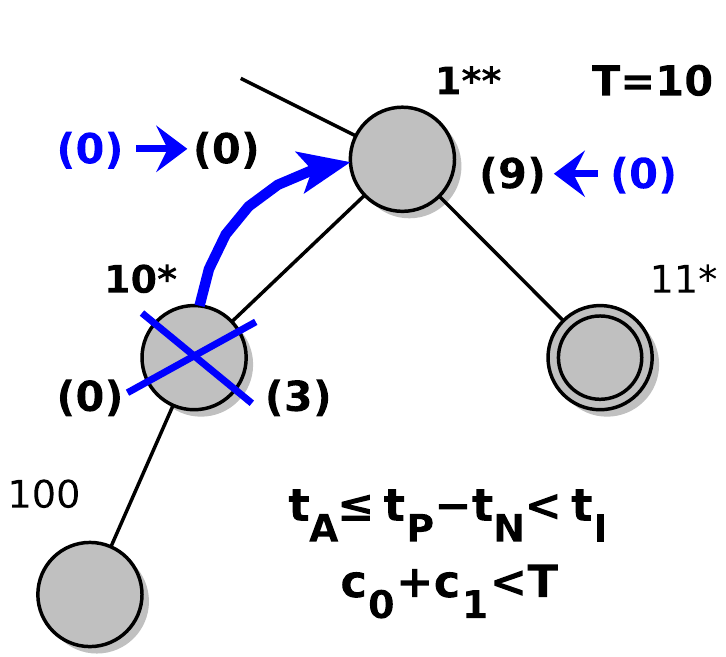}\vspace*{0.5em}
                \caption{Collapsing the node.}
                \label{fig:collapsing-node}
        \end{subfigure}
        \vspace{-0.08in}
        \caption{The core cases of Elastic Trie refinement, assuming the threshold $T=10$. Each node represents a prefix and builds the data structure. Node counters are shown in brackets on the sides.}
        \label{fig:detection-example}
\end{figure*}

\textbf{Invalidating the node.} If the inactive timeout $t_I$ expires ($t_I \le t_P - t_N$), it 
means the prefix node has been inactive for a long time. The values of the counters are outdated and are
not relevant for the detection any more. This can happen when the source prefix stops sending packets for 
a while. Because the detection is built on a packet-driven basis this can not be detected easily in the dataplane.
Thus, the inactive timeout mechanism helps to handle this situation when the packets of the source prefix
start to flow again and when the old values must be invalidated. Figure~\ref{fig:invalidating-node} illustrates
this case. Regardless of the counter values, the tree node is simply removed and the counter values discarded.

\textbf{Expanding the node.} This is the case when both the active and inactive timeouts 
have not expired yet ($t_P - t_N < t_A < t_I$), but one of the node counters (let us assume, for example, 
$c_0$) exceeds the threshold $T$ that the system uses to discriminate heavy prefixes ($c_0 \ge T$). In this case, the
subprefix associated with $c_0$ can be (optionally) reported to the controller as HH but not as HHH yet. 
Figure~\ref{fig:expanding-node} depicts this case: the data structure automatically starts the refinement
of the prefix (\texttt{10*}) by creating a new child node (\texttt{100}) corresponding to $c_0$. According to
the definition of HHH, the original $c_0$ must be set to zero to remove the contribution of the newly created
descendant prefix. Since, we also do not have any records for the newly created child yet, the new node will
have its timestamp set to the current packet timestamp and both its counters set to zero.

\textbf{Keeping the node.} This is the case when the inactive timeout $t_I$ has not expired, 
but the active timeout $t_A$ has expired ($t_A \le t_P - t_N < t_I$), and the sum of both counters exceeds 
the threshold $T$ ($c_0 + c_1 \ge T$), but none of the counters contributes enough to reach the threshold
individually ($c_0 < T$; $c_1 < T$). This 
case is shown in Figure~\ref{fig:keeping-node}. When such a condition happens, the prefix corresponding 
to the node (\texttt{11*}) is a HHH, because it exceeds the threshold $T$ and none of its children contributes 
enough to exceed to threshold individually. The prefix then is reported to the controller, its timestamp 
updated with the packet timestamp value and the counters are reset to prepare the node for the evaluation 
in next time interval.

\textbf{Collapsing the node.} If the inactive timeout $t_I$ has not expired yet, the active 
timeout $t_A$ has expired ($t_A \le t_P - t_N < t_I$) and the sum of both counters does not exceed the
threshold $T$ ($c_0 + c_1 < T$), the node can be collapsed. This case is depicted in Figure~\ref{fig:collapsing-node}.
The node (\texttt{10*}) is removed from the tree structure, and it is replaced by the nearest parent.
Both the counters of the parent node (\texttt{1**}) are zeroed and the timestamp is set to the current 
packet timestamp. Also note the difference between collapsing and invalidation of the node. In the case of
invalidation the nearest parent is not reinserted or renewed.

\textbf{Updating the node counter.} This is the only action which is performed when both the
active and inactive timeouts have not expired yet ($t_P - t_N < t_A < t_I$) and none of the node counters
exceed the threshold $T$ ($c_0 < T$; $c_1 < T$). In this scenario, the node counter corresponding to the
packet subprefix is updated\footnote{Depending on the trie configuration, the counters might carry information
about bytes or number of packets.} and the trie structure does not change. Note the counter is also
updated after other actions when the node is kept, expanded or collapsed. In these cases the newly created node
or the nearest parent node counters are updated instead of the current node counter.

\subsection{Elastic Trie Prototype in P4}

\begin{figure}[t]\centering
    \includegraphics[width=.8\linewidth]{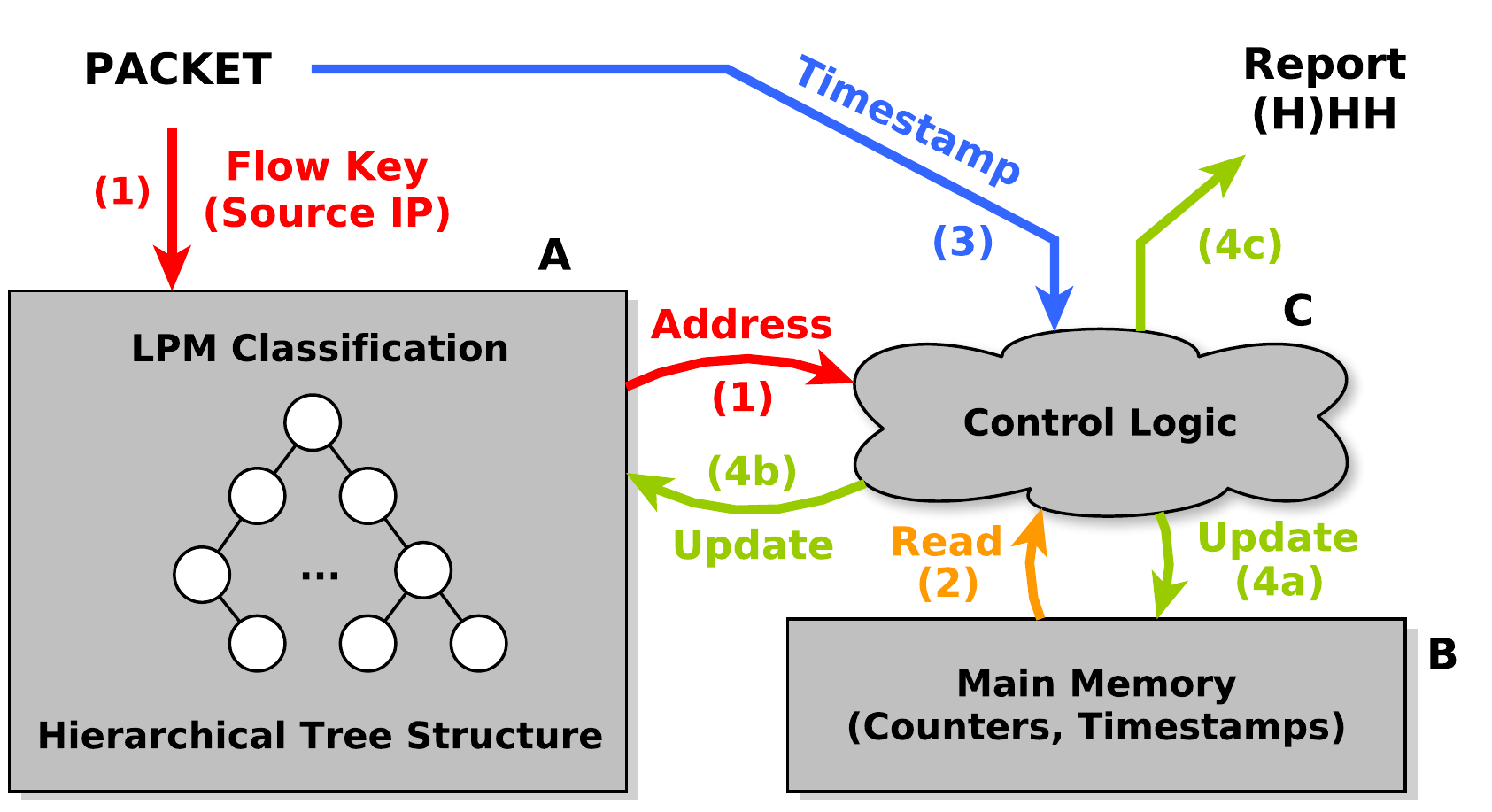}
    \vspace{-0.1in}
    \caption{Elastic Trie dataplane architecture.}
    \label{fig:dataplane-architecture}
\end{figure}

This section discusses the implementation of Elastic Trie on programmable hardware, using 
the P4$_{16}$ specification version 1.0.0~\cite{p4_16}. Figure~\ref{fig:dataplane-architecture} 
depicts a high-level view of the architecture and illustrates the operations performed 
for each incoming packet. The structure is organized around three main building blocks:
(A) the LPM classification stage, (B) the main memory used to gather traffic statistics 
alongside related timestamps and (C) the control logic to dynamically adjust the hierarchical 
data structure and to report (partial) results of the detection to the external controller.

Each incoming packet is first parsed to extract the desired flow key, i.e., source IP\footnote{While
Elastic Trie is oblivious with respect to the specific packet field used as flow key, the 
source IP address is commonly used for the HH and HHHs detection.}. Then, the hierarchical 
tree structure is accessed to find the LPM (step~1). The result of this stage is an address 
that is used to access the main memory, where the data structure of the associated node 
is stored (step~2). The reported values are thus compared 
with the packet timestamp (step~3) and the appropriate operation is computed (step~4) 
following the Elastic Trie specifications described in the previous section. Specifically,
the comparison can trigger an update of the main memory (step~4a), an update of the LPM 
classification scheme (step~4b), or a push notification to the external controller (step~4c).
In the following, we provide a more detailed description of the mapping between the three
main building blocks and $P4_{16}$ match-action constructs.

\begin{figure}[t]\centering
    \includegraphics[width=\columnwidth]{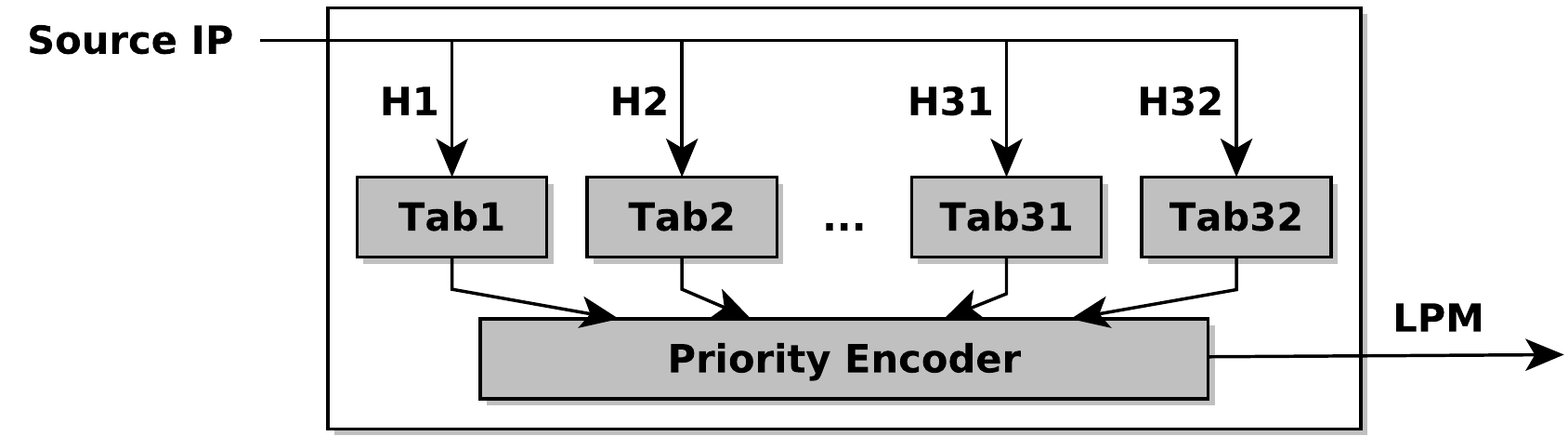}
    \vspace{-0.24in}
    \caption{The LPM classification stage in P4.}
    \label{fig:lpm}
\end{figure}

\textbf{LPM classification stage.} Although P4 offers built-in match tables supporting 
LPM, we could not utilize them for implementation of the trie structure, since the 
latest P4 specification does not support modifications of these tables directly from 
the dataplane, even though some targets like FPGAs may support it. As this feature is essential 
for our architecture, we opted for a custom LPM implementation. We use a hash table for 
each prefix length (Figure~\ref{fig:lpm}), thus requiring 32 hash tables to support each IPv4 prefix\footnote{Using 
less hash tables and supporting only a subset of prefixes comes at the cost of node 
complexity. Indeed, each node needs to store a counter for each associated subprefix. 
This means that if we use only hash tables for just the prefixes $\backslash$8, $\backslash$16, 
$\backslash$24 and $\backslash$32, we need 
to construct nodes with 256 counters each.}. Each hash table is implemented as a register 
array. Upon packet arrival, all the hash tables are read in parallel, by hashing the 
associated prefix of the flow key. We use \texttt{hash} extern API with \texttt{CRC32}
as an algorithm to generate hash values to access the registers. 
Hash tables referring to short prefix values usually require less memory, as they 
need to store information for a smaller number of results. Thus, depending on the amount 
of memory allocated to each hash table, we use a direct access based only on the prefix 
value itself (so called \texttt{IDENTITY} hash algorithm in P4 API) for some of the
shortest prefix tables. 
Each individual hash table lookup result can then be represented as a single bit value, 
\texttt{1} (found), \texttt{0} (not found) respectively. Using bitwise operators we put 
these bits together to form a bitvector, which serves as an input key into a static ternary 
match table that implements a structure similar to a priority encoder.

\textbf{Main memory access mechanic stage.} The hash value of the resulting LPM is used 
as address to access a register array that stores the required node structure information
for that specific prefix, i.e., two packet counters and a timestamp. We use 32-bit wide packet counters
and 48-bit wide timestamp as it is available in the packet metadata structure in P4. To detect hash
collisions in our implementation of LPM classification stage in P4, we further extended the node data
structure with a up to 32-bit wide flow key field (IPv4 prefix). Note that we
do not need to store the prefix length because we use a separate hash table for each length. Thus, the
size of each node structure is 144\,bits (112\,bits for the node and 32\,bits for the IP address). Then, 
in the case of a hash collision, the nearest shorter prefix node is used instead of the reported LPM.

\textbf{Control logic.}
This last stage compares the packet timestamp with the node timestamp
and applies the logic described in Section~\ref{algo}. The node collapse or expansion is 
performed by updating the appropriate hash table storing the specific prefix that needs 
to be adjusted, while the push-based mechanic is implemented by generating a packet 
digest (\texttt{digest} extern object in P4 API) containing the IP prefix detected as HH
or HHH alongside its node information such as the sum of the counters and the timestamp.
The controller does not directly participate on the trie refinement and receives only
generated messages. Thus, for further evaluation of the control logic and using an
available API of P4 behavioral model \cite{bmv2} we also implemented a lightweight
command line tool in Python to receive and dump reported (H)HH prefixes.

\subsection{Enhanced algorithm}\label{ssec:enhancements}
In this part we introduce three further extensions of the basic algorithm described
in the previous section. First, we address the support for the detection of other 
network events, i.e., superspreaders and network traffic changes. Second, we provide
an optimization that accelerates the trie building process, thus improving the reactiveness 
of our solution.

\textbf{Superspreaders detection.}
As introduced in section~\ref{sec:problem_definition}, a superspreader is 
defined to be a host that contacts at least a given number of distinct destinations
over a short time period. Thus, to enable such a detection, it is important to keep 
track of the number of destinations contacted by each source prefix. To address this
challenge we used a standard Bloom filter~\cite{bloom1970}, which is a memory-efficient
probabilistic data structure commonly used to test for set membership. Specifically,
we deployed the filter in parallel to the main memory to test if a packet belongs to
a new unique flow or not. The key to index the filter consists of the source IP prefix
looked up during LPM classification phase and destination IP address of the packet.
The control logic to dynamically adjust the hierarchical structure is kept the same 
as in the basic algorithm, however, a test-and-set operation on the filter is performed 
for each incoming packet and the appropriate node counters are updated only if a new 
unique flow is detected. This change to the architecture allows the detection of (hierarchical) 
superspreaders using the Elastic Trie data structure. Moreover, the Bloom filter can be 
easily implemented in P4 as a bit array placed in a register and a set of $k$ hash 
functions.

\textbf{Change detection.}
The common way to detect chan\-ges in the traffic patterns is based on deriving a model of 
normal behavior based on the past traffic history and looking for significant changes in 
short-term behavior that are inconsistent with the model itself. One of the desired 
properties for Elastic Trie was to be historical network trend aware (Section~\ref{ssec:desired_prop}).
Indeed, by tracking the number of nodes expanded or collapsed over an active timeout 
interval $t_A$, it is possible to spot sudden changes. To enable such a detection, we 
added a global timestamp register and an integer counter which is incremented and decremented 
when any node of the tree is expanded or collapsed, respectively. When the traffic is steady, 
the number of nodes expanded and collapsed should be similar and the value of the counter 
should vary around zero. Otherwise, if the value of the counter is above or below a specified 
threshold, it denotes a significant change in short-term traffic behavior which is reported to 
the controller using a digest message.

\textbf{Variable active timeout.}
The starting condition for our data structure is associated to a trie composed by a single node, 
corresponding with the zero-length prefix \texttt{*}. Depending on the packet flow, the trie 
is then built to focus on the heavy prefixes. Although the refinement process, as explained in 
Section~\ref{algo}, does not depend on the selected active timeout, the node evaluation\footnote{ 
The process of deciding if a specific prefix is a HH or HHH.}, with the potential reporting to 
the controller, does. This means that in the worst case scenario a full IP address can be reported after 
$32 \times t_A$ seconds: the upper bound for 
building the tree from the root to the lowest level. To mitigate this, we propose a variable active timeout mechanism which
sets different timeout intervals and corresponding thresholds for nodes of different prefix
length, i.e., smaller timeout and threshold for shorter prefixes and vice versa. In the P4 implementation
we can use separate configuration registers of active timeout and threshold for each level of the
tree depending on the prefix length. 

\section{Evaluation}
Following a common practice adopted for the evaluation of programmable dataplane
solutions~\cite{liu16,sivaraman17}, we implemented a C++ simulation model of
the Elastic Trie algorithm to assess our approach against real traffic traces
from an ISP backbone and a datacenter network. Additionally, using the behavioral model
available for P4 switches~\cite{bmv2}, we also verified the correctness of our
P4 prototype by comparing its results against the outputs
generated by the C++ simulation model.

In this section, we first describe our setup and we evaluate the trade-offs of
the Elastic Trie data structure. Then, we discuss its detection accuracy against
the supported network events (hierarchical heavy hitters, superspreaders and
traffic changes) when varying memory occupancy, type of the input traffic and
data structure configuration parameters. Finally, we conclude by comparing it
with prior related solutions.

\textbf{Traces.} For the ISP backbone test case, we used four one-hour
packet traces from CAIDA~\cite{Caida:2009,Caida:2016} recorded from 10\,Gbps
links in San Jose and Chicago in 2009 and 2016, respectively. All CAIDA traces
are distributed in one minute chunks, and each chunk contains on
average 30M packets with around 840K unique IP addresses. As for datacenter network test
case, we used the publicly available traces from 2009~\cite{benson-traces}.
These are 65 and 160 minutes long and contain about 20M and 100M packets, respectively,
both with around 5.5K unique IP addresses. Unfortunately, we could not use
the newer datacenter traces from the Facebook Network Analytics Data Sharing
program~\cite{fb}, as they were collected using sampling, which makes
them inappropriate for the type of tests needed in this paper.

\textbf{Setup.} Following common practices from past research
efforts~\cite{liu16,sivaraman17} we set the fixed active timeout $t_A$ to 20 seconds
(measurement reporting time) and the inactive timeout $t_I$ to 5 minutes.
The threshold $T$, used to discriminate the prefixes that are "large enough",
has been set to be 1\%, 5\% and 10\% of the maximum amount of traffic (packets
or flows).
As for the variable active timeout behavior (discussed in Section~\ref{ssec:enhancements}),
when adopted, we set it differently for each trie level. The set $f_y(x)$ of functions
(\ref{eq:f_y(x)}) specifies the value of the timeout for each of the
trie level $x$. The coefficient $y$ indicates the number of levels not being
affected. For example, a value of 16 for $y$ means that the first half
of the trie is built using variable active timeout and the second one
with a fixed timeout.

\begin{equation}\label{eq:f_y(x)}
f_y(x) = \begin{cases}
\frac{y}{32-x}\,t_A, & \text{if}\ \frac{y}{32-x}<1; \\
\;\;\;\;t_A,\;\;\;\; & \text{otherwise}
\end{cases}
\end{equation}

\begin{figure*}[t!]\centering
    \begin{subfigure}{0.33\textwidth}\centering
        \includegraphics[width=\textwidth]{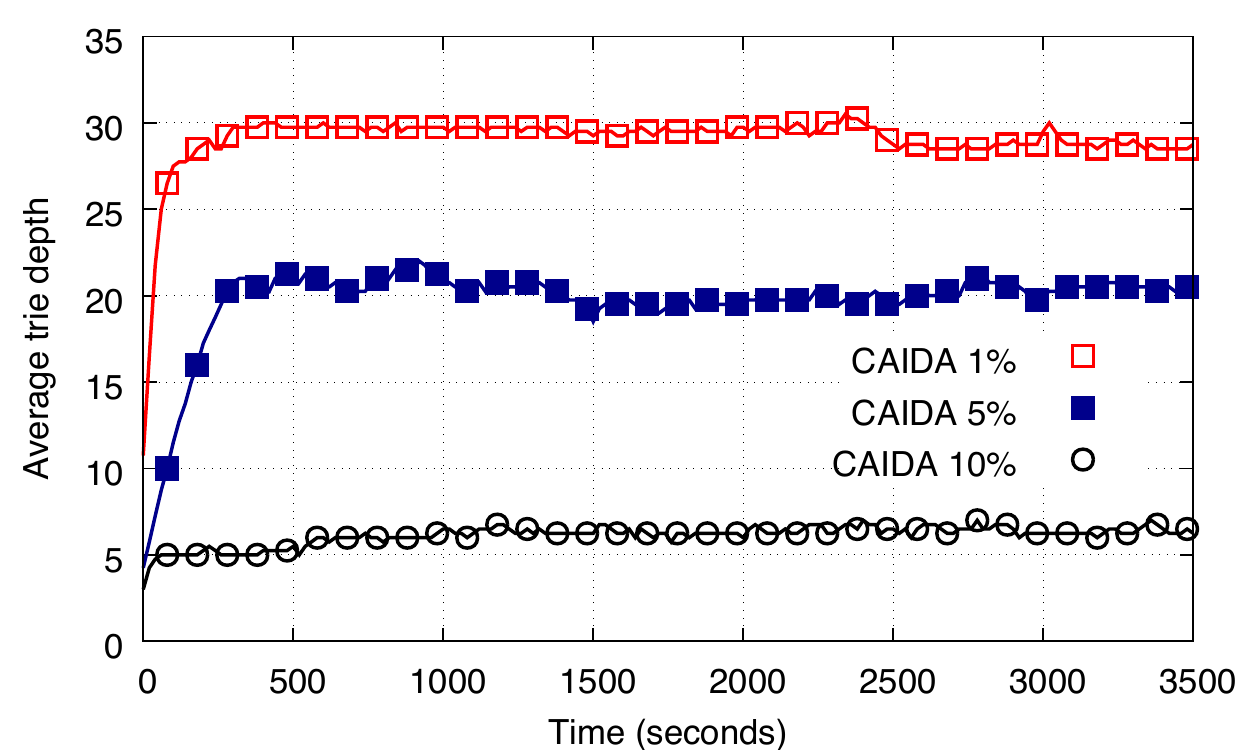}
        \includegraphics[width=\textwidth]{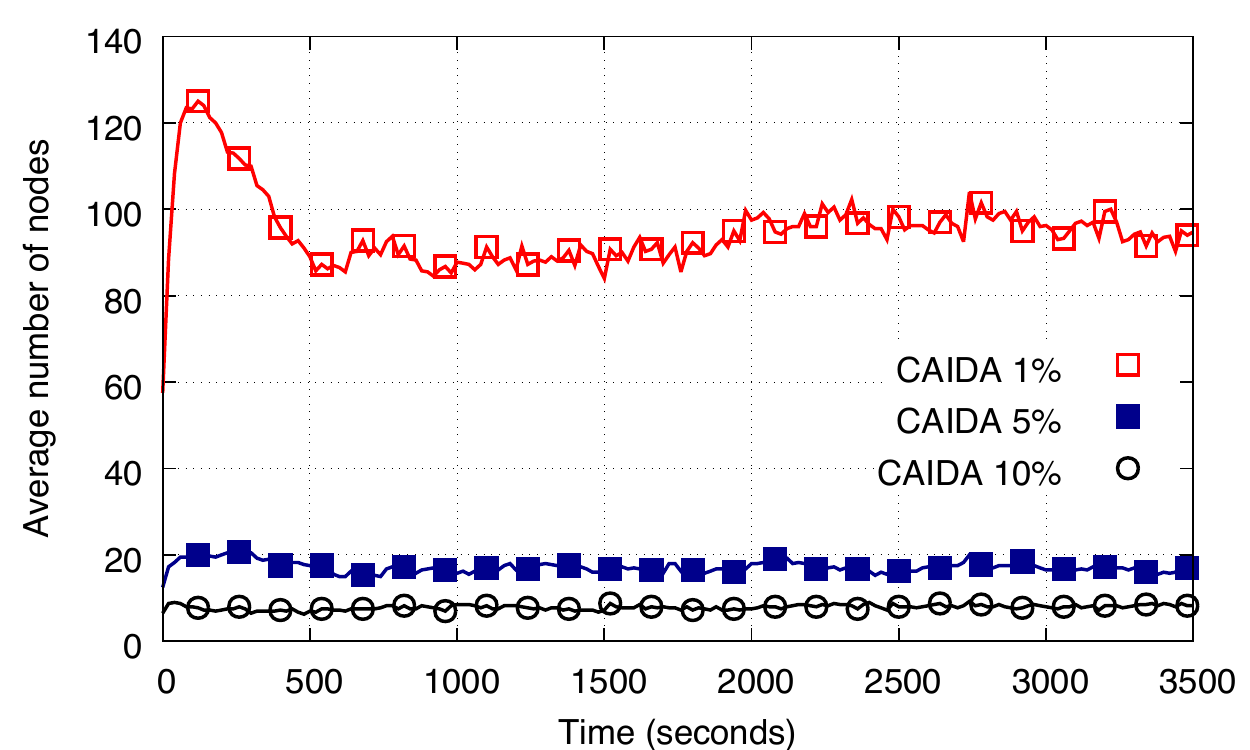}
        \captionsetup{justification=centering}\vspace{-0.2in}
        \caption{Fixed timeout $t_A$\,\\ in a ISP scenario.}
        \label{fig:et-depth-caida-fix}
    \end{subfigure}\begin{subfigure}{0.33\textwidth}\centering
        \includegraphics[width=\textwidth]{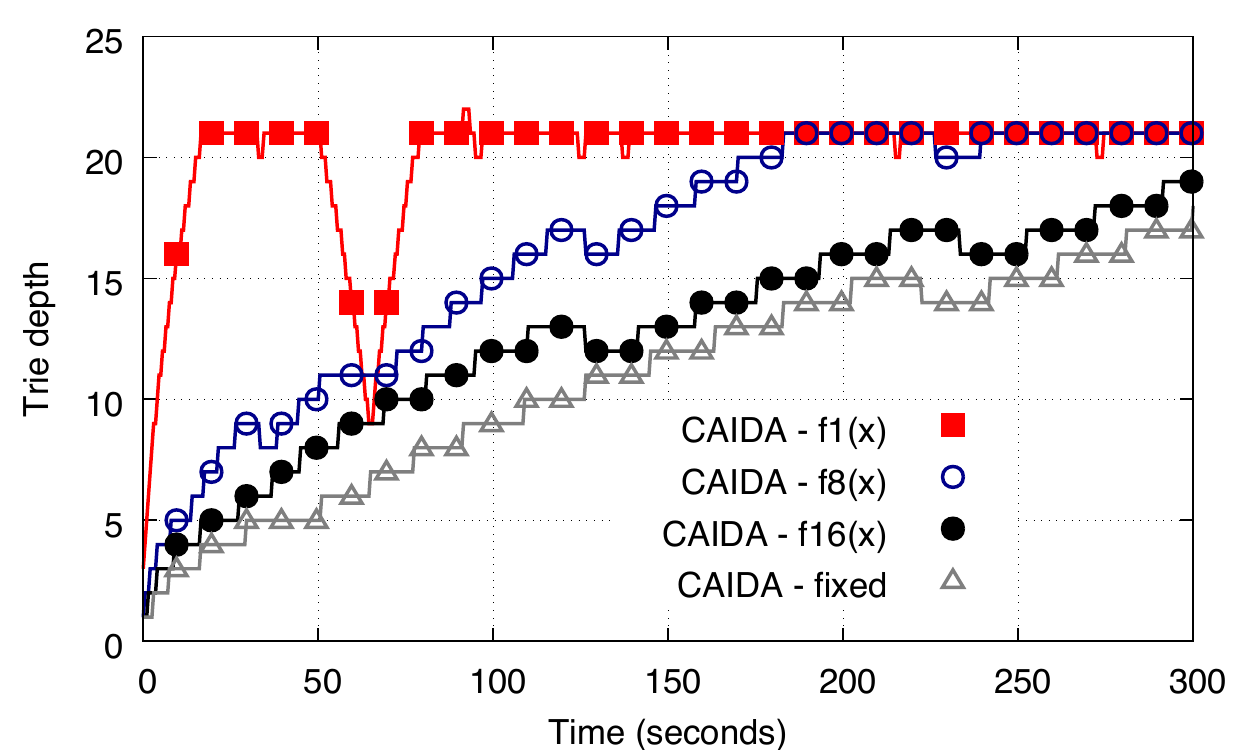}
        \includegraphics[width=\textwidth]{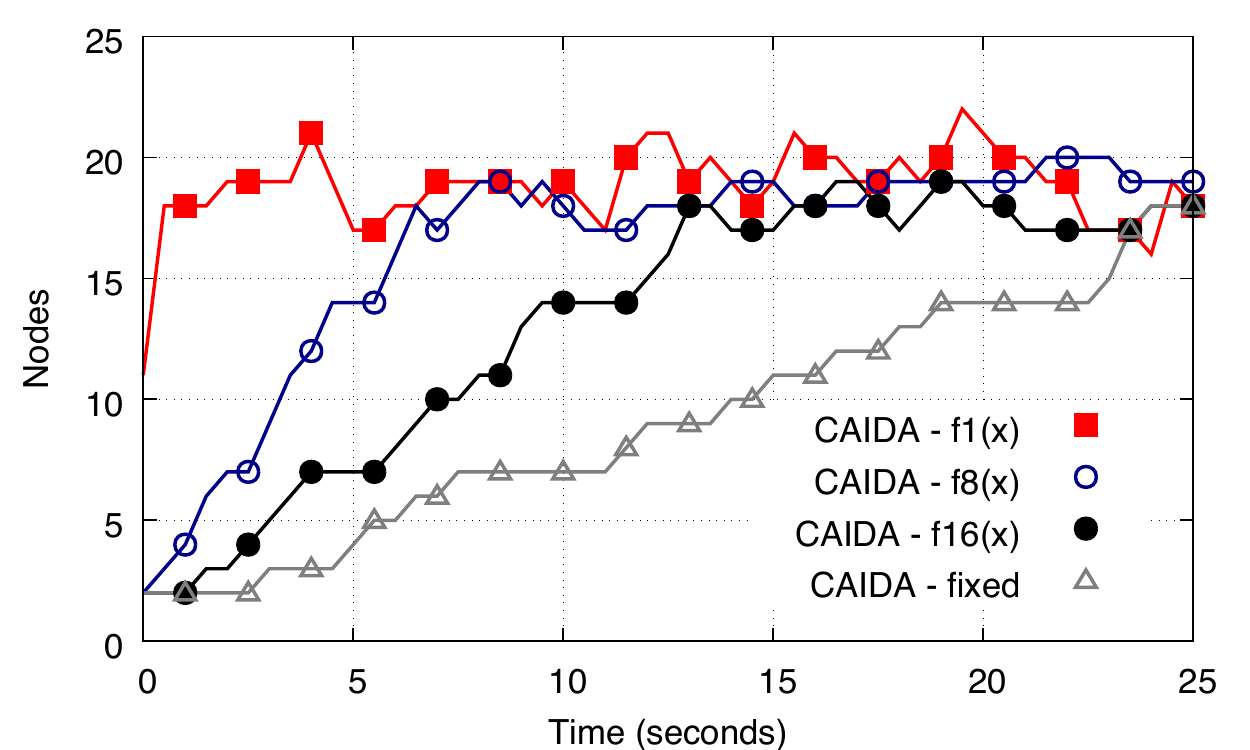}
        \captionsetup{justification=centering}\vspace{-0.2in}
        \caption{Variable timeout $t_A$\,\\ in a ISP scenario.}
        \label{fig:et-depth-caida-var-zoom}
    \end{subfigure}\begin{subfigure}{0.33\textwidth}\centering
        \includegraphics[width=\textwidth]{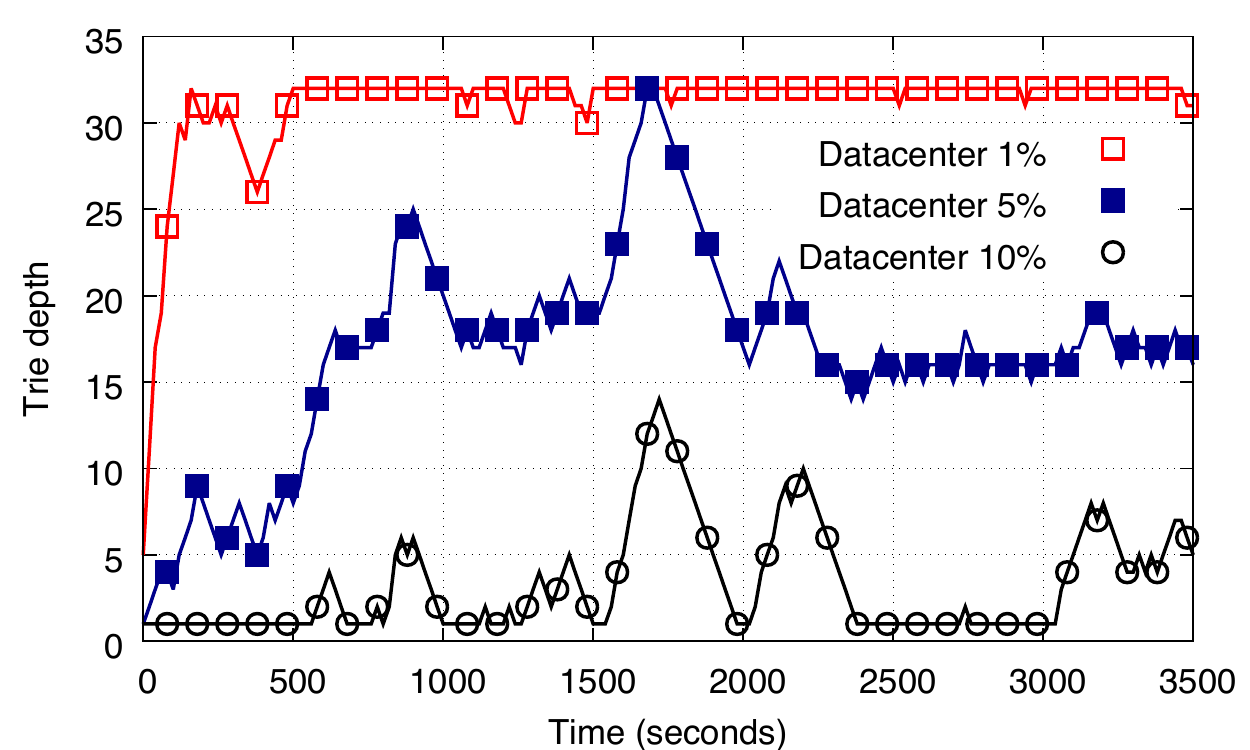}
        \includegraphics[width=\textwidth]{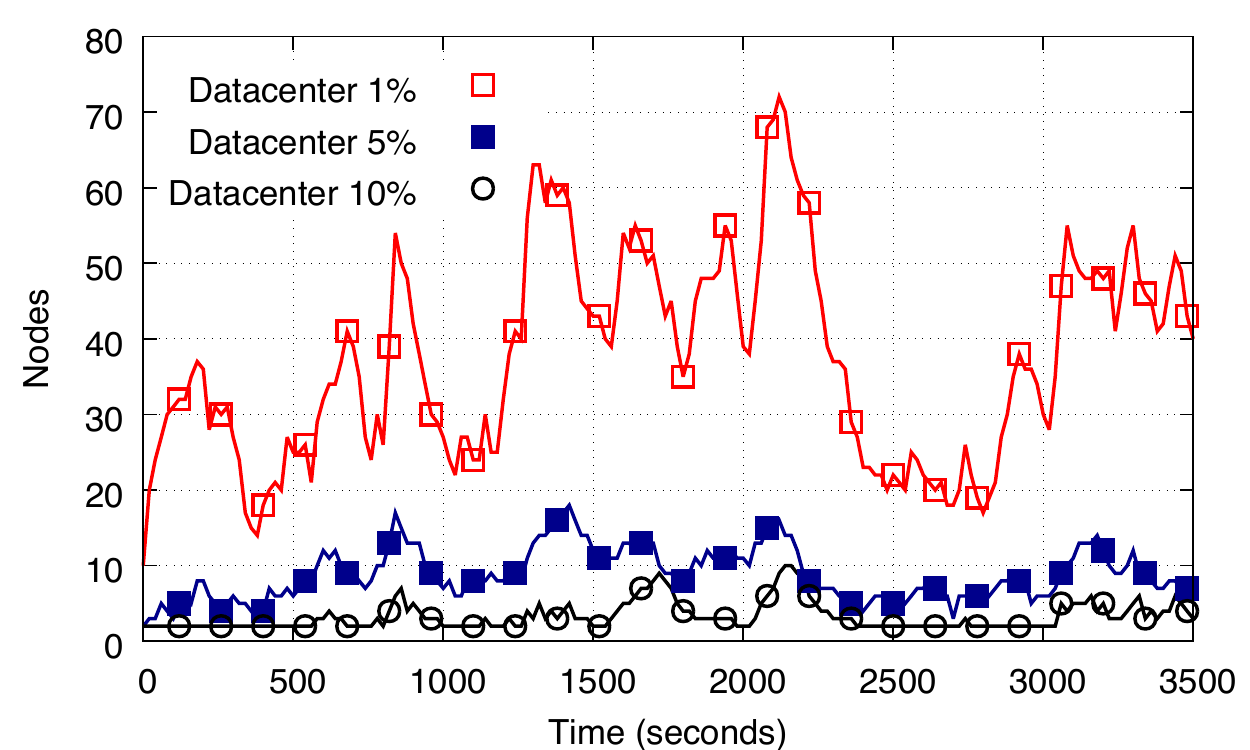}
        \captionsetup{justification=centering}\vspace{-0.2in}
        \caption{Fixed timeout $t_A$\,\\ in datacenter scenario.}
        \label{fig:et-depth-dc-fix}
    \end{subfigure}
    \vspace{-0.1in}
    \caption{Trie depth and number of nodes varying threshold, timeout behavior and type of traffic.}
    \label{fig:et-depth}
\end{figure*}

This set of functions allows to have smaller timeouts for shorter prefixes, thus
enabling a fine-grained control over the reporting time and the trie building
capabilities. The shorter the timeouts,
the smaller the amount of traffic needed to start the process of trie building,
since the threshold $T$ is fixed.

\textbf{Metrics.}
To better understand the dynamics of the proposed data structure, we evaluated the number
of nodes and the trie depth varying a number of configuration parameters.
Then, to estimate its network event detection capabilities, we used two common metrics~\cite{cormode08,jose11}:
recall and precision. Recall~(\ref{eq:recall}) is defined as the number of real events
reported over the total ground truth events happened. In contrast, precision~($\ref{eq:precision}$)
represents the total ground truth events happened over the total reported.
Specifically, the recall and precision are the complements to the number of false negatives $f_N$
and false positives $f_P$: the higher the recall the smaller the false negatives rate, while the higher
the precision the smaller the false positives rate.
\begin{equation}\label{eq:recall}
recall = \tfrac{\text{\emph{real-events-reported}}}{\text{\emph{ground-truth-events-happened}}} = 1-f_N
\end{equation}
\begin{equation}\label{eq:precision}
precision = \tfrac{\text{\emph{ground-truth-events-happened}}}{\text{\emph{events-reported}}} = 1-f_P
\end{equation}
Unless otherwise stated, the recall and precision are always indicated as the
average over the chunks of the traces.

\subsection{Data structure properties}

Figures~\ref{fig:et-depth-caida-fix} show Elastic Trie average depth and average number of nodes
over time for CAIDA traces varying the threshold. The threshold $T$ was set to 1\%, 5\% and 10\%
of the amount of traffic in terms of packets. The depth and number of nodes are as expected proportional to the
selected threshold: the lower the threshold, the larger the depth and the number of nodes,
since more prefixes are detected as heavy.
It is also possible to see the learning phase of the trie at the beginning of the trace, when the trie
has to build up from the less specified prefix. After this phase, the trie reaches a steady state
that reflects the current traffic behavior. Figures~\ref{fig:et-depth-caida-var-zoom} offer, for the
threshold of 5\%, a more detailed view on the learning phase and compare the impact of variable
active timeout for different functions. Using a variable active timeout mechanism, we can speed up
the learning phase by 93\%, going from 300 seconds needed for fixed timeout to 20 seconds
needed using the most aggressive function $f_1(x)$.
As a counterpart, very aggressive functions are much more sensitive to traffic patterns, resulting
in potential fluctuations of the trie. The last Figures~\ref{fig:et-depth-dc-fix} show the behavior
of Elastic Trie in a datacenter environment. In contrast with ISP traces, datacenter traces are much
more bursty, directly influencing the behavior of the trie.

\subsection{HHH detection}

In this section we first present the theoretical HHH detection capabilities,
then our implementation driven results. The former case does not take into
account the impact of implementation details such as amount of available
memory or potential hash collisions. This allows us to get an understanding
about the behavior of our solution in the best case scenario. The latter
takes into account limitations in memory availability, as well as potential 
hash collisions that might happen during the classification stage.
This allows us to get an understanding of the trade-offs between memory 
and detection results.

\begin{figure}[t!]\centering
    \begin{subfigure}{\columnwidth}\centering
        \includegraphics[height=3.6cm,trim={6 2 16 8},clip]{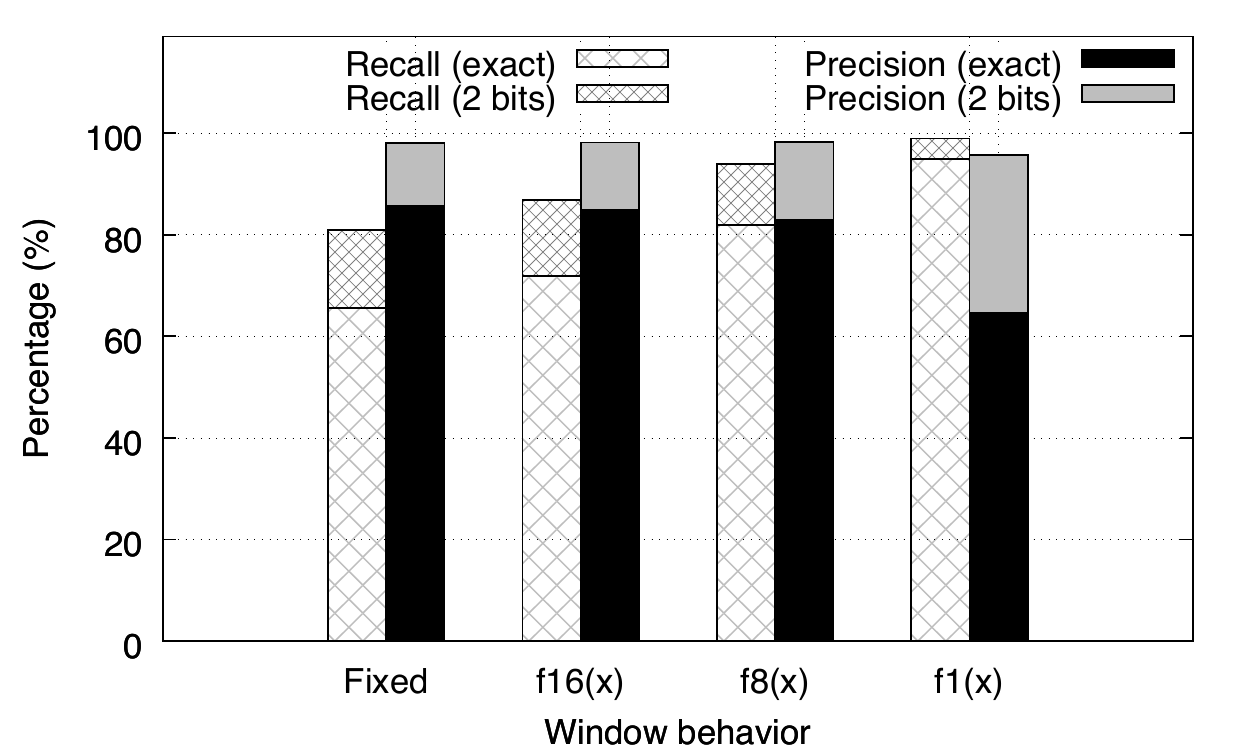}
        \vspace{-0.06in}
        \caption{ISP scenario}
        \label{fig:hhh_detection_caida}
    \end{subfigure}\vspace{0.06in}
    \begin{subfigure}{\columnwidth}\centering
        \includegraphics[height=3.6cm,trim={6 2 16 8},clip]{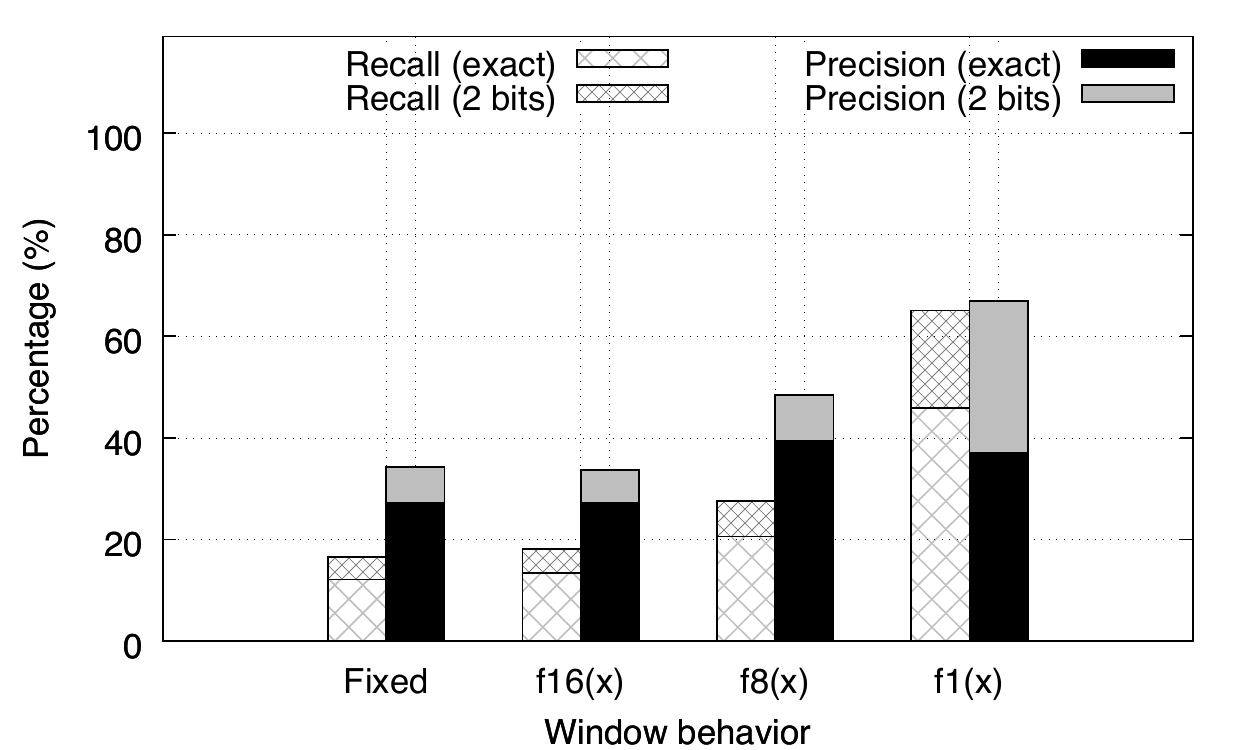}
        \vspace{-0.06in}
        \caption{Datacenter scenario}
        \label{fig:hhh_detection_dc}
    \end{subfigure}
    \vspace{-0.1in}        
    \caption{HHH detection capabilities varying active timeout behavior and type of traffic.}
    \label{fig:hhh_detection}
\end{figure}

\textbf{Theoretical results.} Figure~\ref{fig:hhh_detection_caida} and~\ref{fig:hhh_detection_dc}
show the HHH detection capabilities in a ISP and a datacenter scenario,
respectively. We used a threshold of 5\%.

Since the basic behavior of Elastic Trie is to build a trie that focuses on the prefixes that
account for a large share of the traffic, sometimes it might happen that the system is not
quick enough in finalizing the building process when the prefix needs to be reported. Due to this,
we define two ways of comparing the prefixes detected: exact prefix comparison and relaxed
comparison, where we accept as a valid result a 2 bit coarser grained version of the prefix.
The figures show that the accuracy with exact prefix detection is lower than its
2 bit coarser grained prefix version. Overall, both recall and precision are always between
90\% and 100\%. The effect of variable active timeout can be seen in Figure~\ref{fig:hhh_detection_caida}.
When using a more aggressive variable timeout the recall increases, leading to a smaller false negative
rate. In contrast, the precision decreases causing higher false positives rate. This is a direct
effect of smaller active timeouts that lead the system to detect more prefixes. Using different
functions for variable active timeout, it is then possible to fine tune the trade-off
between recall and precision.

In a datacenter scenario, as shown in Figures~\ref{fig:hhh_detection_dc}, results are less accurate.
This is caused by the bursty nature of datacenter traffic, which means it is
more difficult for the trie to build up in time. It is then clear that our solution is more
suitable for an ISP scenario.

\begin{figure*}[t!]\centering
    \begin{minipage}{.32\textwidth}\centering
        \begin{subfigure}[tc]{\textwidth}
                \includegraphics[height=3.6cm,trim={17.5 0.5 16.5 9},clip]{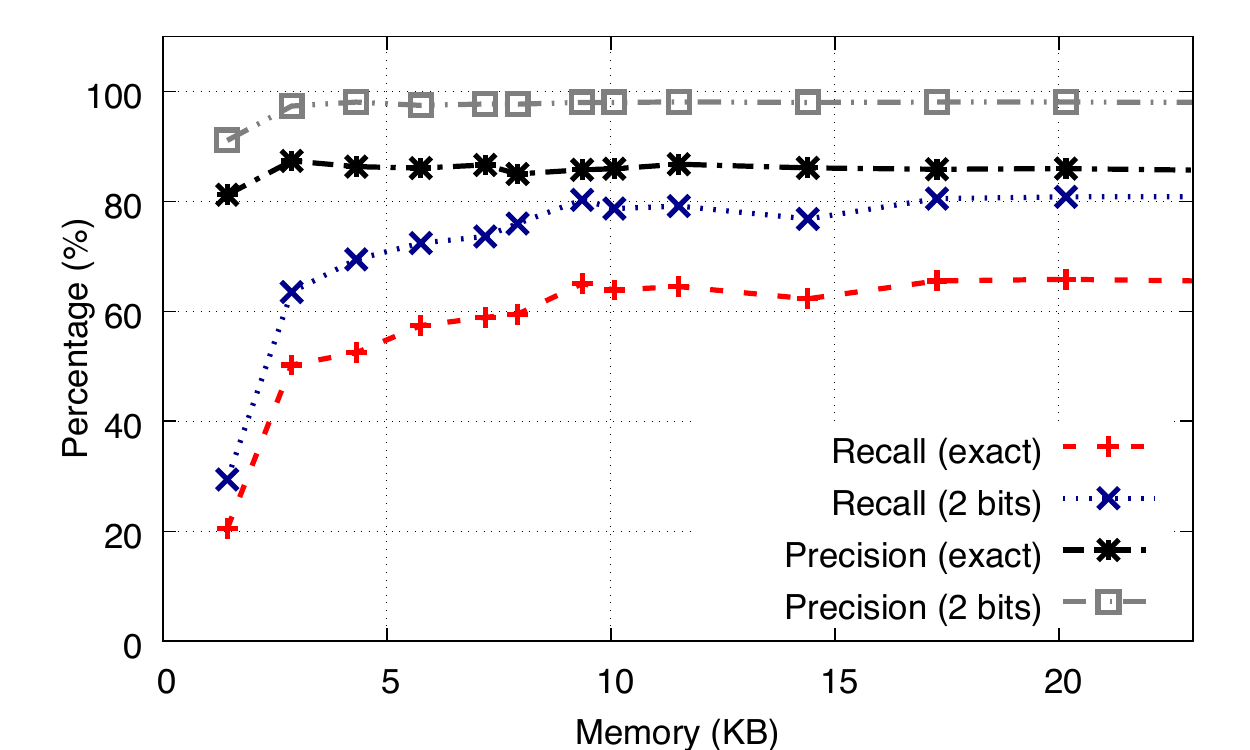}
                \vspace{-0.22in}\caption{Fixed timeout $t_A$}
                \label{fig:hhh_tradeoff_fix}
        \end{subfigure}\vspace{0.08in}
        \begin{subfigure}[tc]{\textwidth}
                \includegraphics[height=3.6cm,trim={17.5 0.5 16.5 9},clip]{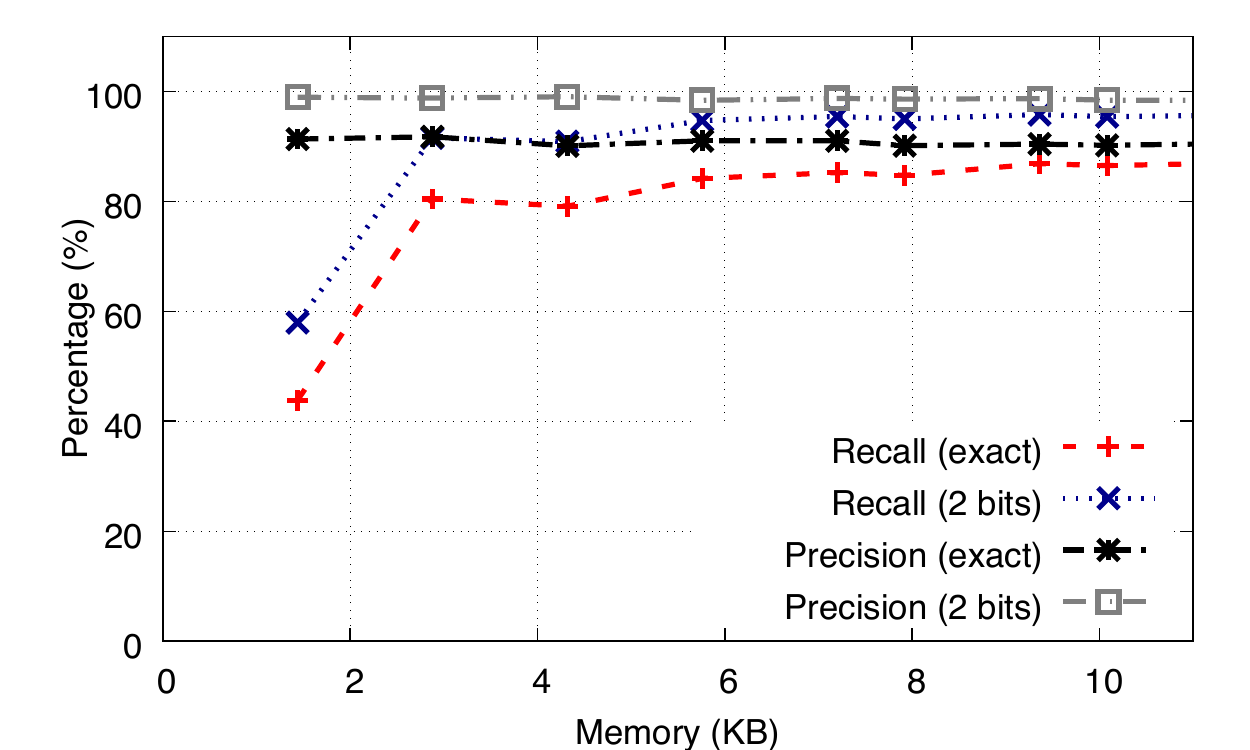}
                \vspace{-0.22in}\caption{Variable timeout $t_A$}
                \label{fig:hhh_tradeoff_var}
        \end{subfigure}
        \vspace{-0.1in}
        \caption{HHH detection capabilities varying memory occupancy and active timeout behavior.}
        \label{fig:hhh_tradeoff}
   \end{minipage}\hfill\begin{minipage}{.64\textwidth}\centering
        \begin{subfigure}[tr]{0.521\columnwidth}
                \includegraphics[height=3.61cm,trim={7 2 17 9},clip]{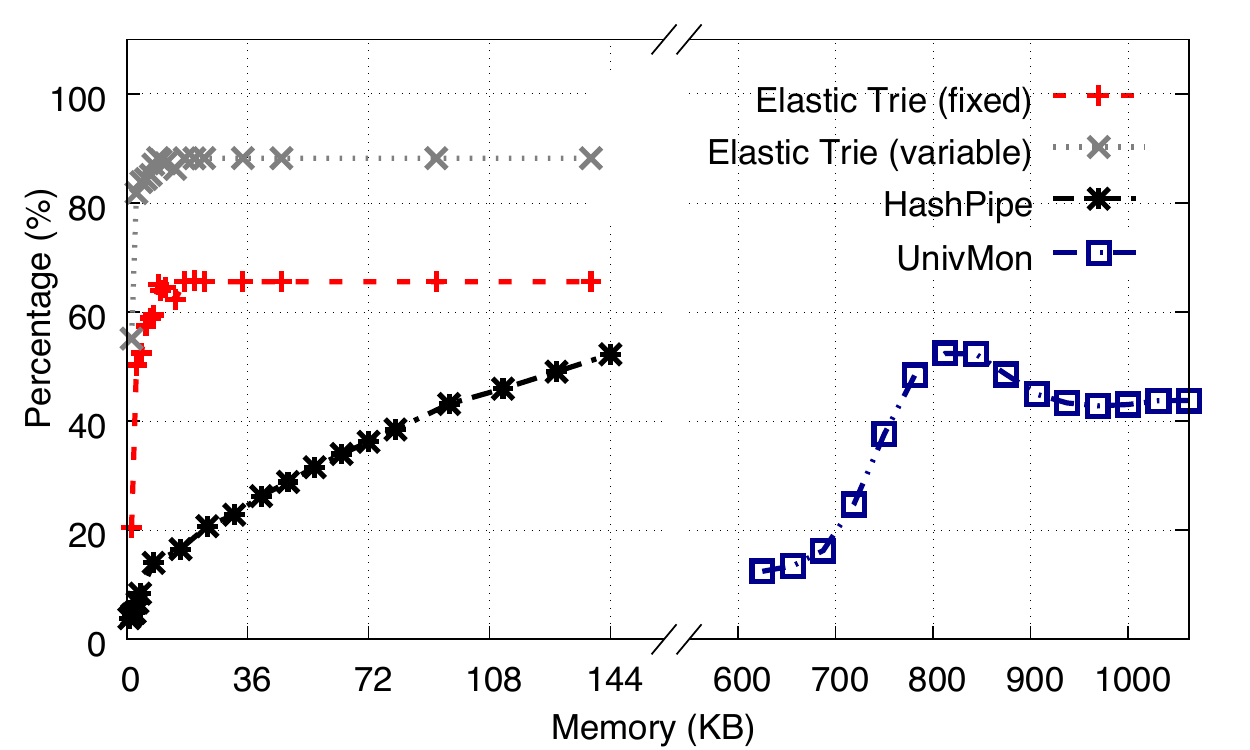}
                \vspace{-0.22in}\caption{Recall (exact)}
                \label{fig:recall0}
        \end{subfigure}\begin{subfigure}[tl]{0.479\columnwidth}
                \includegraphics[height=3.61cm,trim={35.5 2 16 9},clip]{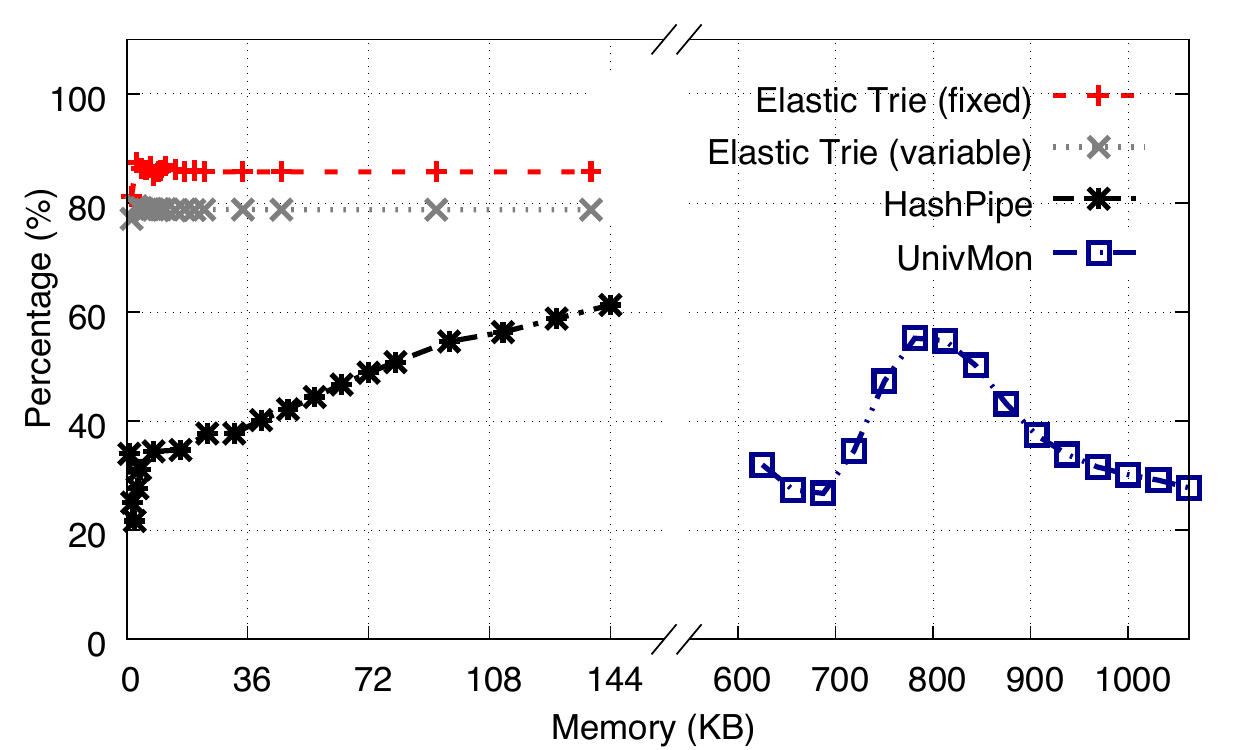}
                \vspace{-0.22in}\caption{Precision (exact)}
                \label{fig:precision0}
        \end{subfigure}\vspace{0.08in}
        \begin{subfigure}[tr]{0.521\columnwidth}
                \includegraphics[height=3.61cm,trim={7 2 17 9},clip]{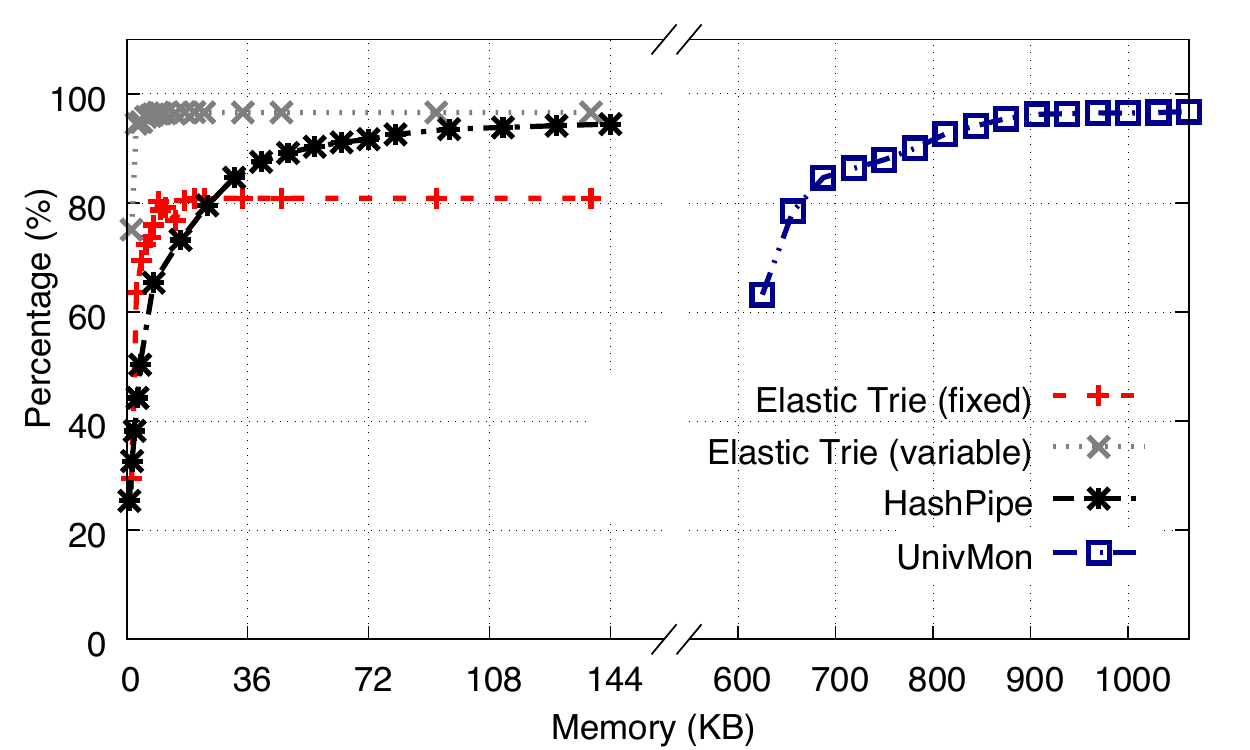}
                \vspace{-0.22in}\caption{Recall (2 bits)}
                \label{fig:recall2}
        \end{subfigure}\begin{subfigure}[tl]{0.479\columnwidth}
                \includegraphics[height=3.61cm,trim={35.5 2 16 9},clip]{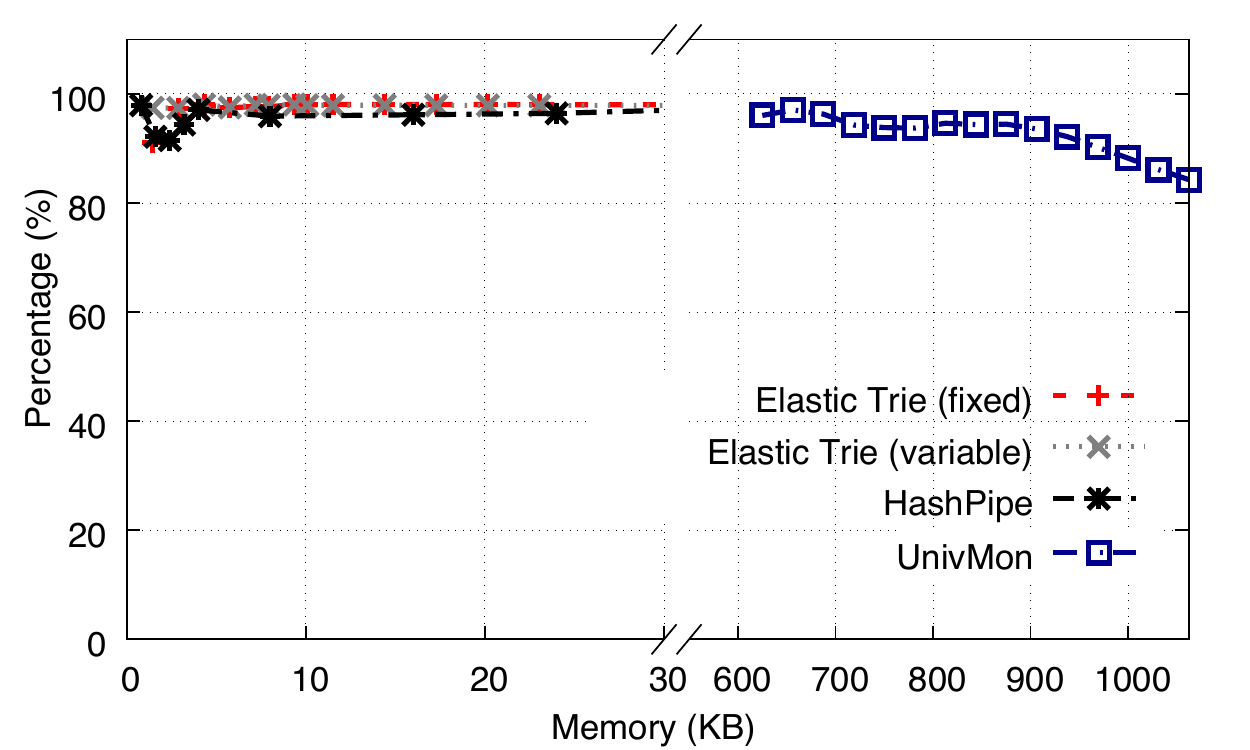}
                \vspace{-0.22in}\caption{Precision (2 bits)}
                \label{fig:precision2}
        \end{subfigure}
        \vspace{-0.1in}
        \caption{Comparison between Elastic Trie, UnivMon, and Hashpipe of Hierarchical Heavy Hitter detection capabilities in ISP scenario.\\}
        \label{fig:comparisons}  
\end{minipage}
\end{figure*}

\textbf{Implementation driven results.} We assess the impact of the amount of available
memory over the recall and precision. We find that our solution can successfully detect, with approximately 65\%
recall and 85\% precision, the exact HHH prefix using a fixed active timeout and less than
20KB (Figure~\ref{fig:hhh_tradeoff_fix}).
If a coarser grained prefix is used, which is less precise by only two bits, then
the recall jumps to 80\% and the precision to 98\%. Again, this is the consequence
of the nature of the data structure: it might happen that it does not have enough time
to build properly.

Using a variable timeout (Figure~\ref{fig:hhh_tradeoff_var}) results improve sensibly. In this
case, it is possible to detect the exact HHH prefix with 85\% recall and 90\% precision with
less than 8KB. Moreover, if a 2 bit coarser grained HHH prefix is used,
the recall jumps to 95\% and the precision to 98\%. Increasing the available memory does
not significantly improve the detection capabilities of the system, because it is theoretically
bounded by the ability of the trie to react and build up according to the input traffic
patterns.

In Figure~\ref{fig:comparisons}, we compare the HHH detection capabilities
of Elastic Trie against related prior programmable dataplane based solutions: UnivMon~\cite{liu16}
and HashPipe~\cite{sivaraman17}.
UnivMon and HashPipe use an alternative definition for HH detection, named the ``top-$k$ problem''.
Instead of reporting prefixes that are larger than a given threshold, they report the top-$k$
sources, no matter the amount of traffic they are actually sending. To perform a fair comparison,
and align their results with the one produced by our system (which follows the classic HHH
definition), we aggregated their output addresses into prefixes and considered only the ones
that carry traffic beyond the fixed threshold $T$. Figures~\ref{fig:recall0} and  \ref{fig:precision0} show the results using an exact prefix comparison. HashPipe needs a much lower amount of memory
($\sim$144KB) than UnivMon ($\sim$800KB) to reach recall and precision around 50-60\%. In contrast,
Elastic Trie significantly outperforms both of the solutions.
The inaccuracies spotted in UnivMon and HashPipe can be partially related to the different
definition of heavy hitters being used. This is also confirmed by the results obtained when
a coarser grained prefix is used (Figures~\ref{fig:recall2} and \ref{fig:precision2}). Nevertheless, the
memory requirements of the three solutions represent a fair comparison metric. HashPipe and Elastic
Trie have the same memory requirements, but HashPipe can only detect Heavy Hitters, while our solution
adds more network events. UniMon is not restricted to a single network event, but requires 90\% more
memory to work.

\subsection{Superspreader detection}

As in the HHH case, we first introduce the theoretical results (without taking into account
available memory or hash collisions). Then, we show the trade-offs between memory occupancy and
superspreader detection capabilities.

\begin{figure}[t!]\centering
    \begin{subfigure}[tc]{\columnwidth}\centering
        \includegraphics[height=3.6cm,trim={6 2 16 8},clip]{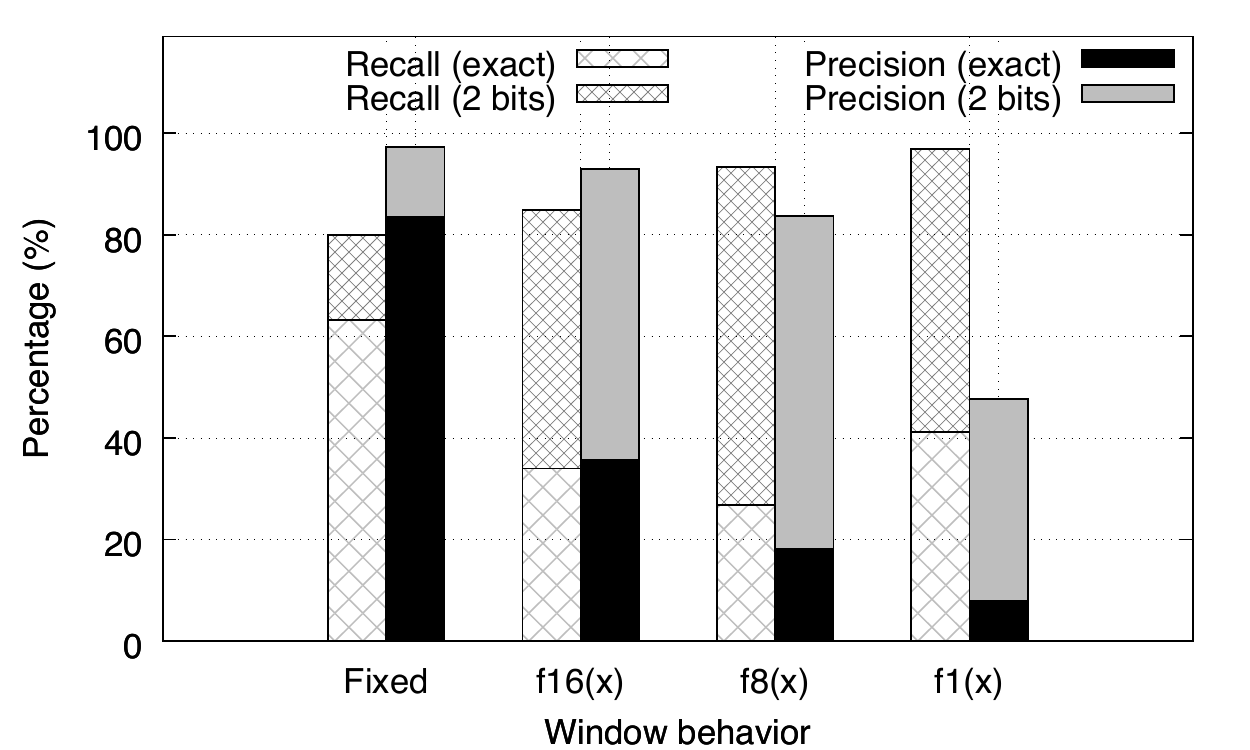}
        \vspace{-0.06in}
        \caption{Variable timeout $t_A$}
        \label{fig:ss_detection}
    \end{subfigure}\vspace{0.06in}
    \begin{subfigure}[tc]{\columnwidth}\centering
        \includegraphics[height=3.6cm,trim={6 2 16 8},clip]{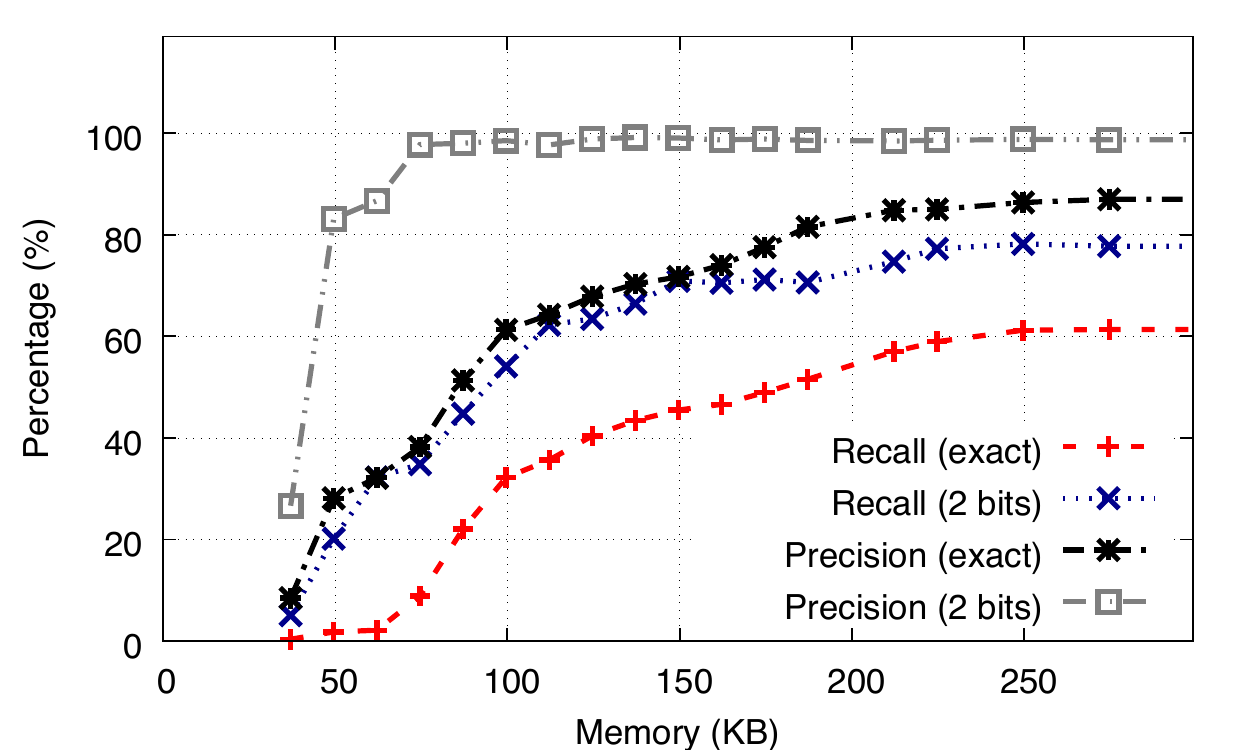}
        \vspace{-0.06in}
        \caption{Memory occupancy}
        \label{fig:ss_tradeoff}
    \end{subfigure}
    \vspace{-0.1in}
    \caption{Superspreader detection capabilities.}
    \label{fig:ss}
\end{figure}

\textbf{Theoretical results.} Figure~\ref{fig:ss_detection} shows the theoretical superspreader detection recall and precision
capabilities for CAIDA traces varying the active timeout behavior. In contrast to the same
evaluation for HHH detection, Elastic Trie superspreader detection, using variable timeout, is
less good at detecting the exact prefix length than when using a fixed timeout.
In this case, it is clearer that the trie cannot build in time, as both recall and precision grow sensibly
when we use a 2 bit coarser grained superspreader prefix. Overall, for fixed active timeout the
detection capabilities are still good, as both recall and precision are around 80\% and 95\%.

\textbf{Implementation driven results.} In Figure~\ref{fig:ss_tradeoff} we show the impact of available memory over the detection
capabilities, taking into account our P4 implementation. For this test, we used a fixed
active timeout, fixed 25KB of memory for the allocation of the prefix trie structure, and
we varied the Bloom filter occupancy. We find that superspreaders can be successfully detected with
approximately 60\% recall and 85\% precision and 78\% recall and 98\% precision when a 2 bit
coarser grained prefix is used, respectively, with less than 250KB of allocated memory.

\subsection{Change detection}

\begin{figure*}[t!]\centering
    \begin{subfigure}{0.33\textwidth}\centering
        \includegraphics[width=\textwidth]{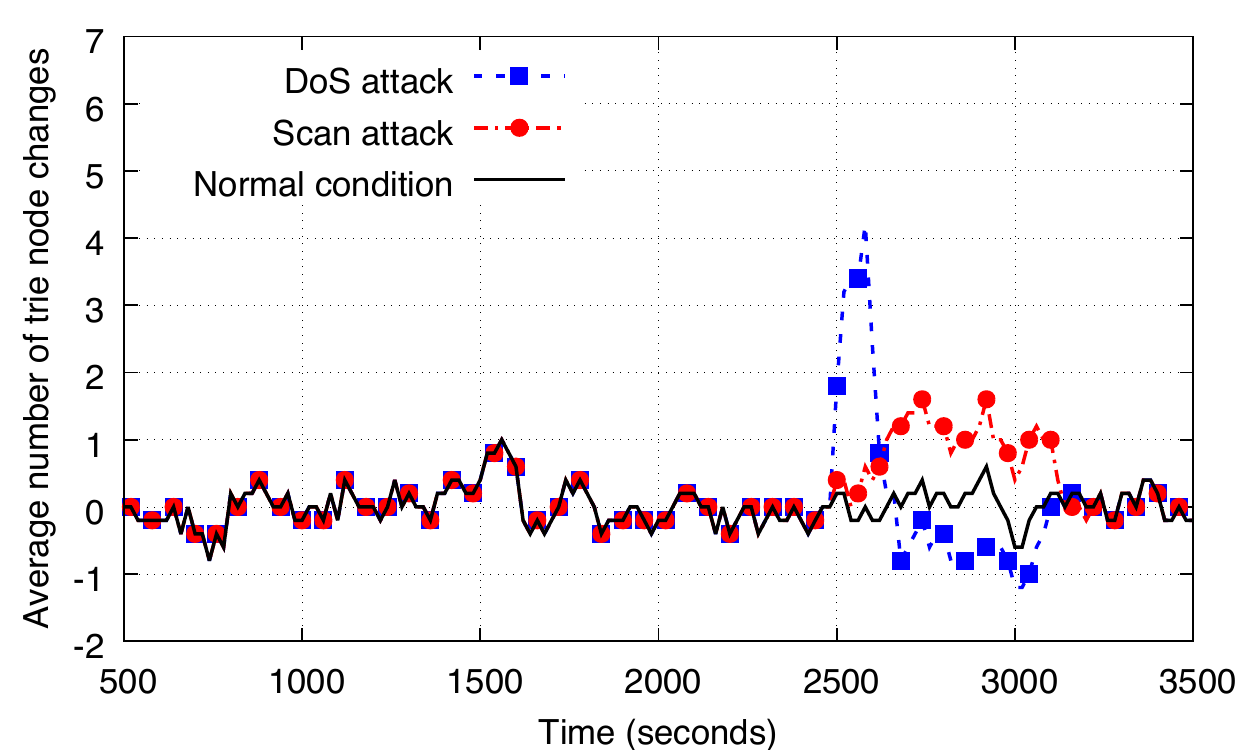}
        \captionsetup{justification=centering}
        \vspace{-0.2in}\caption{Fixed active timeout,\\ using (H)HH detection.}
        \label{fig:cd-fix-pp}
    \end{subfigure}~
    \begin{subfigure}{0.33\textwidth}\centering
        \includegraphics[width=\textwidth]{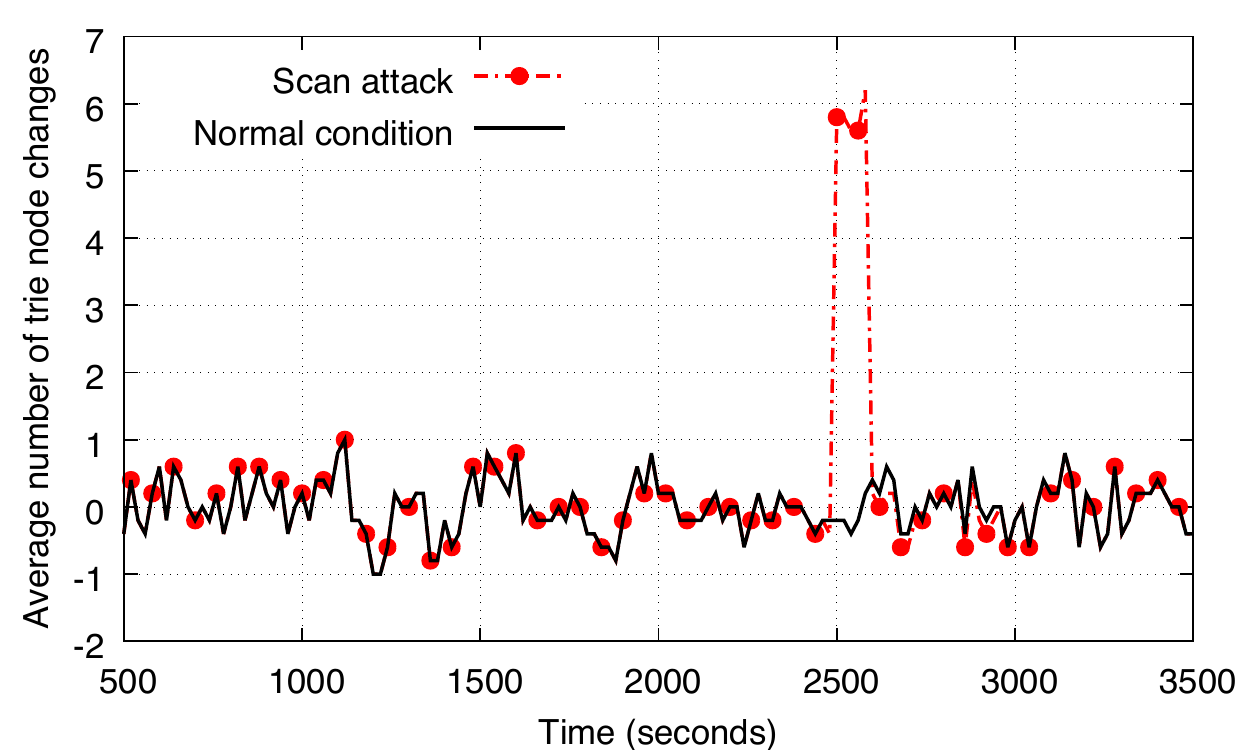}
        \captionsetup{justification=centering}
        \vspace{-0.2in}\caption{Fixed active timeout,\\ using superspreaders.}
        \label{fig:cd-fix-fp}
    \end{subfigure}~
    \begin{subfigure}{0.33\textwidth}\centering
        \includegraphics[width=\textwidth]{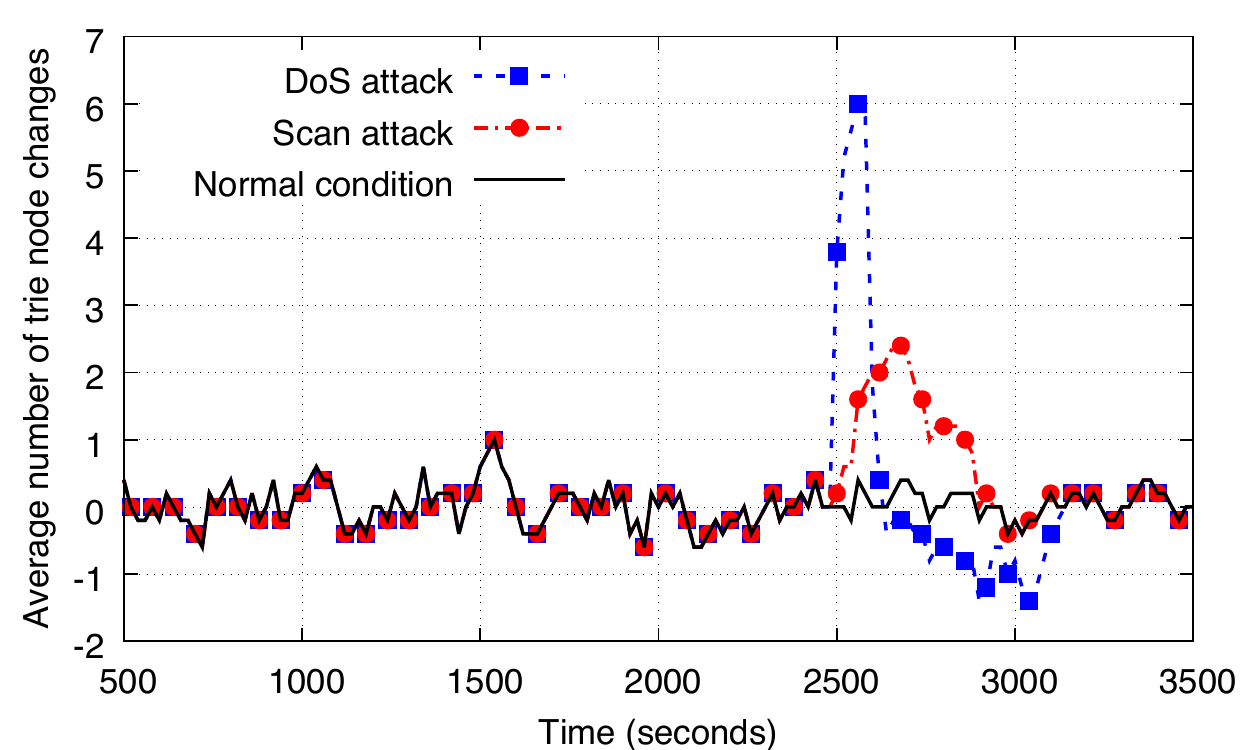}
        \captionsetup{justification=centering}
        \vspace{-0.2in}\caption{Variable active timeout,\\ using (H)HH detection.}
        \label{fig:cd-var-pp}
    \end{subfigure}
    \vspace{-0.1in}
    \caption{Change detection capabilities varying active timeout and trie building behavior.}
    \label{fig:cd}
\end{figure*}

To demonstrate traffic change detection capabilities of Elastic Trie structure, we artificially
injected network traffic simulating DoS attack and scanning into one of the CAIDA traces. The
attack has been emulated after 2500 seconds since the beginning of the trace. DoS and scanning
are two type of attacks that can potentially change traffic patterns. At the same time, they
are also pretty different: while a DoS is typically a source that sends a huge amount of traffic
to a designated victim, the scan is a source contacting many random destinations.
Figure~\ref{fig:cd-fix-pp} shows the time on the x-axis and a moving average of trie changes (difference
between number of expanded and collapsed nodes) on the y-axis. Note that tree is built based on HHH
detection using a fixed active timeout $t_A=20$ seconds. In the figure we can distinctly see differences
during normal conditions and the state under DoS attack or scan. After a learning phase,
our data structure can be used to verify if the input traffic patterns suddenly change.
Figure~\ref{fig:cd-fix-fp} shows the same situation but from a different perspective. Now the trie
is built on top of the superspreader detection. In this case, the DoS attack is not detected at
all, because it represents a communication with only one distinct destination. On the other hand, the
scan, as a typical case of superpreader, is much more significant now.
The last Figure~\ref{fig:cd-var-pp} shows the situation based again on HHH detection, but using the variable
active timeout mechanism. Due to accelerated trie construction, there are many more changes in the
trie over a short time period. This allows to highlight further even small changes in the traffic patterns,
as shown when comparing the scan behavior for Figure~\ref{fig:cd-fix-pp} and Figure~\ref{fig:cd-var-pp}.

\section{Related Work}
SDN-based monitoring solutions which rely on statistics retrieval from 
switches~\cite{OpenTM, yu13, PayLess} might suffer from limited visibility.
As demonstrated in Section~\ref{sec:motivation}, this can be a very expensive
process, that can overload either the controller or the switch itself.
In contrast, our solution reports to the controller the network events of 
interest as soon as they happen, without the need of a central controller.
To have flexibility in identifying the interesting flows, iterative refinement 
of monitored flows can be used, which can be costly for the control channel, 
since it requires flow updates to zoom in the traffic of interest~\cite{jose11, zhang13, moshref2014}.
While our architecture also relies on iterative refinement when building the 
trie to focus on the flows of interest, it does so in the data plane, 
with no direct intervention from the control plane. 
Algorithms that use iterative refinement of flows to determine heavy hitters  
and anomalies were presented in~\cite{yuan2007} and~\cite{khan2011}, respectively, but
they were targeted for custom measurement platforms (not match-action type 
architecture).

More recently, a number of monitoring frameworks leveraging P4 programmability have
been developed~\cite{li16, liu16, sivaraman17, popescu17}. FlowRadar~\cite{li16} keeps track of all the flows  
in the network and their counters, and exports this information periodically to
a remote collector, which decodes it and uses it for various monitoring applications targeted 
to datacenters. Our aim is to not keep track of all the flows 
in the network as FlowRadar does, but to be able to efficiently detect network
events related to high-volume traffic clusters from within the dataplane.
UnivMon~\cite{liu16} uses a general sketch
in the dataplane to keep track of the flows, which offers information for several monitoring
applications, and is exported at fixed time intervals to the control plane.
HashPipe~\cite{sivaraman17} determines the top-k heavy hitters in the dataplane, and exports them
at fixed time intervals. Our work informs the controller as soon as the 
considered network events take place, without having to wait for the end of the time interval. 
Sonata~\cite{sonata}, which proposes a query interface for network telemetry, 
uses sketches in the dataplane, and zooms-out the network traffic of interest 
by refining the network query, starting from the finest level. The query refinement
is done by the controller, who programs the new query plan on the target in each iteration, 
while in our case, the refinement is done directly in the dataplane. While~\cite{popescu17} presents
a solution for determining hierarchical heavy hitters directly from the dataplane, their solution 
does not cover other measurements tasks.   


 







\section{Conclusion}
In this paper, we proposed a push-based approach to network monitoring, where 
the dataplane informs the control plane only when specific conditions are met.
To achieve this, we presented a new data structure, Elastic Trie, that enables 
the detection of a number of network events associated with high-volume traffic
clusters  within the dataplane. Our solution has been designed with the constraints 
of emerging programmable switches in mind, as it works in a packet-driven manner, 
and can be implemented using common match-action based architectures such as RMT.

Elastic Trie uses a hash table based prefix tree that grows or collapses to 
focus only on the prefixes that account for a "large enough" share of the traffic. 
This enables the detection of (hierarchical) heavy hitters, and by looking at 
its growing rate it is possible to identify changes in the traffic patterns.
We prototype our solution with P4 and demonstrate its detection capabilities.
Specifically, using simulation on real traffic traces taken from an ISP 
backbone and a datacenter, we showed that Elastic Trie achieves high accuracy
in detecting hierarchical heavy hitters, superpreaders and changes in the
network traffic patterns with the memory constraints imposed by today's switches. 

\label{ConcPage}
\bibliographystyle{abbrv}
\bibliography{paper}

\end{document}